\documentclass[11pt]{article}
\pdfoutput=1
\usepackage{amsmath}
\usepackage{amssymb}
\usepackage{graphicx,bbm,mathrsfs}
\usepackage{nicefrac}
\usepackage{bbm}
\usepackage{geometry}       
\usepackage{jheppub}       

\def\ie{{\it i.e.}}

\newcommand{\be}{\begin{equation}}  
\newcommand{\ee}{\end{equation}}  
\newcommand{\bea}{\begin{eqnarray}}  
\newcommand{\eea}{\end{eqnarray}}

\newcommand\lsim{\mathrel{\rlap{\lower4pt\hbox{\hskip1pt$\sim$}}
    \raise1pt\hbox{$<$}}}
\newcommand\gsim{\mathrel{\rlap{\lower4pt\hbox{\hskip1pt$\sim$}}
    \raise1pt\hbox{$>$}}}

\newcommand{\captionfonts}{\small}

\newcommand{\approptoinn}[2]{\mathrel{\vcenter{
  \offinterlineskip\halign{\hfil$##$\cr
    #1\propto\cr\noalign{\kern2pt}#1\sim\cr\noalign{\kern-2pt}}}}}

\makeatletter  
\long\def\@makecaption#1#2{%
  \vskip\abovecaptionskip
  \sbox\@tempboxa{{\captionfonts #1: #2}}%
  \ifdim \wd\@tempboxa >\hsize
    {\captionfonts #1: #2\par}
  \else
    \hbox to\hsize{\hfil\box\@tempboxa\hfil}%
  \fi
  \vskip\belowcaptionskip}
\makeatother   

\begin{document}

\vspace*{1.2cm}

\begin{center}

\thispagestyle{empty}
{\Large\bf New likelihoods for shape analysis }\\[10mm]

\renewcommand{\thefootnote}{\fnsymbol{footnote}}

{\large Sylvain~Fichet$^{\,a,b}$\footnote{sylvain.fichet@lpsc.in2p3.fr, sylvain.fichet@gmail.com}}\\[10mm]

\addtocounter{footnote}{-1} 

{\it
$^{a}$~ICTP South American Institute for Fundamental Research, Instituto de Fisica Teorica,\\
Sao Paulo State University, Brazil \\
}

{\it
$^{b}$~International Institute of Physics, UFRN, 
Av. Odilon Gomes de Lima, 1722 - Capim~Macio - 59078-400 - Natal-RN, Brazil \\
}

\vspace*{12mm}

{  \bf  Abstract }
\end{center}

We introduce a new kind of likelihood function based on the sequence of  moments of the data distribution. 
Both binned and unbinned data samples are discussed, and the multivariate case is also derived. 
Building on this approach we  lay out the formalism of shape analysis for signal searches. In addition to moment-based likelihoods, standard likelihoods and approximate statistical tests are provided. Enough material is included to make the paper self-contained from the perspective of shape analysis.
We argue that the moment-based likelihoods can advantageously replace unbinned standard likelihoods for the search of non-local signals, by avoiding the step of fitting Monte-Carlo generated distributions. 
This benefit increases with the number of variables simultaneously analyzed.
The moment-based  signal search is exemplified and  tested in various 1D toy models mimicking typical high-energy signal--background configurations. 
Moment-based techniques  should be particularly appropriate for the searches for effective operators at the LHC. 
\\
\\
\textbf{Keywords}: new physics searches; statistical methods

\noindent 

\clearpage
%

\section{Introduction}  \label{se:introduction}

 The search for New Physics is a rather challenging task.  At the quantum level, physical phenomena are described by  probability distributions.  
  The measurements of such quantum observables  typically consists in collecting events whose occurrence in time is described by these probability densities.  
Consider a continuous observable $X$ taking values over a domain $\mathcal{D}$ following a probability  distribution $f_X$.
    The simplest measurement possible   is generally the global event counting over  $\mathcal{D}$. 
 In such case, the event rate  is  proportional to the probability $\int_\mathcal{D}dx\, f_X(x)$. 
 In order to gain further knowledge, the next logical step is to try to learn more about the $f_X$ distribution itself. 
 Typically, the experiment measuring $X$ is then set up in order to divide $\mathcal{D}$ into domains $\mathcal{D}_r$ as small as possible. 
 The event counting in each bin $\mathcal{D}_r$ provides a discrete estimator of $f_X$, and the smaller the bins, the larger the gain of information.

Getting information about $f_X$ from its estimator is  of tremendous importance in various experimental situations. 
For example, the heavy physics possibly lying beyond the Standard Model (SM) can be parametrized by effective operators of higher dimension. 
These operators are suppressed by powers of the new physics mass scale. Their effects might be too tiny to be observed as a deviation from total event rates,  
while they  could instead be spotted inside the kinematic distributions of the observed particles.
A familiar  example is the one of Higgs physics. At the LHC, the first Higgs observables released were the event rates. These measurements can be  translated as constraints   on the Higgs effective operators (see e.g. \cite{Dumont:2013wma}). However, certain degeneracies among operators can be lifted only when considering the shape of  kinematic distributions \cite{Ellis:2014dva}. 

It is clear that the analysis of the shape of $f_X$  is an exercise that can be frequently encountered. 
Given its importance, shape analysis deserves a careful treatment in order to be optimized, both at the statistical and the technical level. 
In this paper, we introduce a new kind of likelihood function based on the truncated  moment sequence  of data distributions.~\footnote{Notice there exists a ``method of moments'' \cite{GMM}.  It is used to characterize a parametric distribution, and thus does not correspond to the topic we treat here.}
 We will argue that the moment-based likelihood can replace standard likelihoods for the searches of a non-local signal. This in turn implies a  simplification of the shape analysis for signal searches,  typically encountered in high-energy physics.
 A review of the standard likelihoods and simplified statistical tests are also included,  such that this paper is self-contained from the perspective of shape analysis. 

We will first outline the standard  shape analysis method and the necessary  statistical basics in Sec.~\ref{se:like_std}. We  introduce the moment-based likelihood in Sec.~\ref{se:like_mom} for both binned and unbinned data,  including the multivariate case. The information content and practical use are also discussed.
We then lay out the formalism of shape analysis for signal searches in Sec.~\ref{se:signal} and display the maximum likelihood estimators used for simplified statistical tests. 
The advantages and limits of the moment-based approach for signal searches are discussed in Sec.~\ref{se:limits}. 
We exemplify in Sec.~\ref{se:toymodels} the moment-based approach on toy-models with shapes typical of high-energy signal searches.  Section~\ref{se:summary} contains the conclusions and outlook.

\section{Notations and standard likelihoods for  shape analysis \label{se:like_std} }

Here we shortly outline basic statistical facts related to shape analysis. The likelihood function\footnote{The so-called likelihood function is actually a distribution.}  $\mathcal{L}$ is the central object that confronts  the hypothetical and observed outcome of an experiment. It is defined as the probability distribution of the observed data taken as a function of the  hypothesis $H$ that one wishes to test. It generally reads ${\cal L}(H) \equiv p({\rm data}|H)$, and is defined up to a multiplicative constant.    For shape analysis the hypothesis $H$ can in particular be a probability density function (PDF)  $f^{\rm }$, or a continuous quantity $\theta$ characterizing a parametric distribution $f^{\rm }_\theta$. 
Any quantity built from the data can in principle be called an ``estimator'', and is usually denoted by a hat. For the number of observed events, that should be  denoted $\hat{n}$  in principle, it is customary to drop the hat when no ambiguity is possible. When doing so, it is identified with its expected value $E[\hat n]=n$.~\footnote{Note $\hat n$ is sometimes called $n_{\rm obs}$ in the literature.}
Through this paper our interest is in the shape of data distributions, and not on the total event rate.  All distributions considered are therefore normalized to one without loss of generality, unless stated otherwise.~\footnote{There is no extra difficulty in including the total event rate in addition to the shape information. 
The combination is described in last subsection of Sec.~\ref{se:signal}.
}

Let $X$ be the  measured observable, $\mathcal{D}$ its domain, and $f_X$ the hypothesized shape of its distribution.
 Assume that an independent sample  of  $X$, denoted  $(X_i)$, is  known with infinite precision. Then, by definition, the likelihood is given by
\be 
\mathcal{L}_{\rm std}\propto\prod_i^n f^{\rm}_X(X_i)\,. \label{eq:Lfull}
\ee
 This likelihood contains the maximum information available from the data.  Any alternative likelihood can contain either  as much or less information than $\mathcal{L}_{\rm std}$.

  In actual measurements, often the values of $X$   cannot be known with infinite precision. This can come in particular from a finite detector resolution, or from an uncertainty in the knowledge of the phenomenon observed.  
 In such case, $\mathcal{D}$ is usually splitted into subdomains ${\cal D}_r$ (i.e. \textit{bins}) such that ${\cal D}=\cup_r{\cal D}_r$. The events are then labelled with respect to the bin ${\cal D}_r$ to which they belong. 
The amount of events $X_i$ in a given bin $\mathcal{D}_r$ is written $\hat{n}_r$, and  is Poisson-distributed.
The  hypothesized  content of the bins is given by $n^{\rm }_r=n \int_{{\cal D}_r}dx\, f^{\rm }_X$. 
  The general binned likelihood then reads
 \be
{\cal L}^{\rm bin}_{\rm std}\propto
\prod_r  \frac{1}{\hat n_r !}n_r^{\hat n_r} e^{-n_r} \label{eq:L_pois}\,.
 \ee

In the small-bin limit for fixed sample size $n$, each bin contains either zero or one event. Moreover,  $f_X$ can be linearised over each bin provided that it is continuous. In this small-bin limit, ${\cal L}^{\rm bin}_{\rm std}$ reduces to $\mathcal{L}_{\rm std}$, the standard likelihood of Eq.~\eqref{eq:Lfull}. This makes clear that $\mathcal{L}_{\rm std}$ is of practical interest, provided that measurements are precise enough to resolve each event separately. 
In the rest of the paper we will denote this limit as the ``unbinned'' case.

%
%
%

Let us now consider the Poisson likelihood  Eq.~\eqref{eq:L_pois} in the limit of large data sample. One notice that the quantity $\hat{n}_r/\hat{n}\equiv \hat{p}_r$ is an estimator of the probability $\int_{{\cal D}_r}dx f_X\equiv p_r$, which is in general biased. \footnote{This is because $\hat{n}_r$ and $\hat{n}$ are correlated by construction, and because of Jensen's inequality $E[1/\hat{n}]\geq 1/E[\hat{n}]$. }
However, for a large sample, the bias asymptotically goes  to zero, 
\be
E\bigg[\frac{\hat{n}_r}{\hat{n}}\bigg]=\frac{E[\hat{n}_r]}{n}\big(1+O(n^{-1})\big)= p_r\big(1+O(n^{-1})\big)\,.
\ee
These $r$ estimators provide thus  a good discrete estimate of $f_X$ for large data sample. 
Moreover, for a large enough number of events in each bin -- typically $\hat{n}_r> O(10)$, the Poisson distribution for each bin tends to the normal (Gaussian) distribution $\mathcal{N}$,~\footnote{The normal distribution is defined such that $X\sim \mathcal{N}(x_0,\sigma^2)$ means that the PDF of $X$ is $f_X(x)=(\sqrt{2\pi}\sigma)^{-1}e^{-(x-x_0)^2/(2\sigma^2)}$. } such that  
 we  approximately have
 \be \hat{n}_r\sim \mathcal{N}(n_r,n_r(1-n_r/n))\,,\quad \textrm{or}\quad 
\hat{p}_r\sim \mathcal{N}(p_r,p_r(1-p_r)/n)\,.
 \label{eq:nr}\ee  
The variance $V[\hat{n}_r]$ can be estimated by $ \hat{n}_r(1-\hat{n}_r/n)\approx \hat{n}_r$.
\footnote{It is customary to assume $\hat{n}_r/n\ll 1$. There is no difficulty in keeping the subleading term if necessary.
}
\footnote{We find the bias  to be $E[\hat{p}_r]=p_r(1-n^{-1}+O(n^{-2}))$ in that case.
}
For large data sample, the likelihood for binned data Eq.~\eqref{eq:L_pois} takes thus the form
\be
{\cal L}^{\rm bin}_{\rm std}\propto \prod_r e^{-(n_r^{\rm} -\hat{n}_r)^2/2 \hat{n}_r}\,, \label{eq:Lbin}
\quad  \textrm{or}\quad  
{\cal L}^{\rm bin}_{\rm std}\propto \prod_r e^{-n\,(p_r^{\rm} -\hat{p}_r)^2/2 \hat{p}_r}\,.
\ee
The well-known feature of this distribution is that the variance for each bin decreases as $1/\hat{n}_r$, \ie~the precision increases as $\sqrt{\hat{n}_r}$.   
The two likelihoods $\cal {L}^{\rm bin}_{\rm std}$, ${\cal L}_{\rm std}$ are --  to the best of our knowledge -- at the center of the most common and well-defined shape analysis.  


Clearly, binning induces a loss of information with respect to the unbinned data. This information loss can be  quantified using the expected Fisher information about a parameter of interest $\theta$, $I_\theta[{\cal L}]=E[ (\partial \log{\cal L}/ \partial \theta )^2 ]$ \cite{lehmann}.
The expected information content of the standard likelihood is 
\be
I_\theta[{\cal L}_{\rm std}]=n\int_\mathcal{D}dx\,\frac{ (\partial f_\theta(x) / \partial\theta)^2 }{f_\theta(x)} \,.
\ee
The information content of the standard binned likelihood is 
\be
I_\theta[{\cal L}^{\rm bin}_{\rm std}]=n\sum_i\,\frac{ (\partial \int_{\mathcal{D}_i }dx\,f_\theta(x) / \partial\theta)^2 }{ \int_{\mathcal{D}_i }dx\, f_\theta(x)} \,.
\ee
It converges to $I_\theta[{\cal L}_{\rm std}]$ when the bins ${\cal D}_i$ are small enough such that both $f_\theta$ and $\partial f_\theta(x) / \partial\theta$ can be linearized over each of them. Otherwise, one has $I_\theta[{\cal L}^{\rm bin}_{\rm std}]< I_\theta[{\cal L}_{\rm std}]$, which quantifies the loss of information due to the binning.
Note that this way of quantifying the information relies on expected values, so that for a given realization of the data sample,
 this provides only a qualitative idea of the information loss.

\section{New approach: the moment-based likelihood}  \label{se:like_mom}

In order to get new insights about shape analysis, let us decompose $f_X$ over an infinite basis of functions $(g_p)$, 
\be
f_X=\sum_p a_p\, g_p\,. \label{eq:decomp}
\ee
Characterizing $f_X$ then amounts to estimate the coefficients $a_p$. One attractive possibility is to use a orthonormal basis of functions for the $(g_p)$. This possibility is discussed in App. \ref{app:ortho}.  
 In the present paper we will focus on an arguably more universal decomposition involving the moments of $f_X$. 
The decomposition is done over the basis of Dirac delta's derivatives $(\delta^{(p)})$,
\be
f_X=\sum_p \frac{m_p}{p!}\,(-1)^p\, \delta^{(p)}(x)\,. 
\ee
The $m_p$ coefficient is the  $p$-th moment of $f_X$, determined by
\be
m_p=\int_\mathcal{D}dx\, f_X(x) x^p\,.   \label{eq:mom_def}
\ee
Note the zero-th order moment corresponds to the overall normalization of $f_X$, and  thus characterizes the global event rate over ${\cal D}$, \ie~ for a non-normalized distribution the estimator of $m_0$ is $\hat{m}_0=\hat{n}$. Here we are interested only in shapes, so the distribution can be normalized to one, and we have $m_0=1$ by definition. 


\subsection{Unbinned data}

From the moment definition Eq.~\eqref{eq:mom_def}, it appears that an asymptotically unbiased estimator for the $p$-th moment $m_p$ is 
\be
\hat{m}_p=\frac{1}{n}\sum_i^n X^p_i  \label{eq:mom_est_unb}\,.
\ee
By the Central Limit Theorem (CLT), every moment estimator $\hat{m}_p$ follows a normal law at large $n$ (se e.g. \cite{Muirhead} for an introduction to CLTs).
Moreover, by construction, all these estimators are evaluated through the same set of data, so that all the $\hat m_p$'s are necessarily strongly correlated. From the CLT, it appears that the vector of moment estimators $(\hat m_p)$ is described by a multivariate normal distribution with mean $(m_p)$, 
 \be(\hat{m}_p)\sim \mathcal{N}( m_p, \Sigma )\,.\ee 
The expected covariance matrix $\Sigma_{pq}\equiv{\rm Cov}[\hat m_p,\hat m_q]$ is found to be
\be
\Sigma_{pq}=\frac{1}{n}\bigg(m_{p+q} - m_pm_q\bigg)\,,
\ee
and an estimator for the covariance  is given by
\be
\hat{\Sigma}_{pq}=\frac{1}{n}\bigg(\hat{m}_{p+q} - \hat{m}_p\hat{m}_q\bigg)\,.
\ee
As a result the moment-based likelihood for  unbinned data  reads
\be
{\cal L}_{\rm mom}=\exp\bigg(-\frac{1}{2}(m_p-\hat{m}_p)^t\, \hat{\Sigma}_{pq}^{-1}\,(m_q-\hat{m}_q)\bigg)\,.
\label{eq:Lmomfull}
\ee

This moment-based likelihood is at the center of our attention in this paper. 
As will be discussed below, in practice the sequence of moments is always truncated.  The truncated moment-based likelihood where the covariance matrix includes the sequence from first to $P$-th moments is denoted by  ${\cal L}_{\textrm{mom},P}$. In this convention the covariance matrix has dimension $P\times P$ and the moment vector has dimension $P/2$.


\subsection{Information content and practical computation}

The way a piece of information is distributed over the moments depends in general on the problem studied. It is for sure that the expected Fisher information becomes complete when the whole sequence of moments is taken into account, that is
\be
I[{\cal L}_{\textrm{ mom},P}]\rightarrow I[{\cal L}_{\rm std}]\,\quad \textrm{for}\quad P\rightarrow \infty\,.
\ee
This remark is however valid only for the expected information. In practice, the set of data is finite, and the behaviour of the observed information needs to be understood carefully.

The amount of data being finite, one must remark that the moment estimation will always break down at some order. Qualitatively speaking, the first moments characterize the global features of the shape (starting with mean, variance, skewness and kurtosis). Going higher in the moment order, one characterizes the local features of the shape.  For a finite $n$, one can intuitively expect that the finite amount of data will in priority provide information on the global features, and at some point the local features of the shape will not be resolved. 

Concretely, from $n$ observed events $X_{1\ldots n}$, only $n$ independent quantities can be constructed. There can be therefore no more than $n$ moments computed from a set of $n$ events. If one insists to include more than $n$ moments in ${\cal L}_{\rm mom}$, the extra moments can be written as a function of the $n$ moments already included. Total correlations are thus present among the set of moments, and result in a \textit{singular} moment covariance matrix $\hat\Sigma$.

There are reasons, however, to expect an (approximately) singular covariance matrix much before the moment number matches the event number. 
For a fixed event number $n$, the moment estimator tends asymptotically to $X_{\rm max}^p/n$ for large $p$, where $X_{\rm max}=\max(X_i)$\,. For  a large enough $p$, one can write
\be
\hat{m}_p=\frac{X^p_{\rm max}}{n}(1+\epsilon_p)\,, \label{eq:mk_lim}
\ee
with $\epsilon_p\ll 1$. Clearly $\epsilon_p$ decreases with $p$. 
If $\epsilon_p$ reaches zero, the moment covariance matrix becomes singular. In practice, for finite $n$, $\epsilon_p$ does not reaches zero. However, as soon as it becomes of order of the computing system precision, the matrix is effectively seen as singular in the numerical computation. 

Whenever the limit described by Eq.~\eqref{eq:mk_lim} happens, the moment estimation is already totally wrong. A proper truncation of the moment sequence should instead happen when $\hat{m}_p$ just begins to deviate from its expected value. But  in practice, this expected value is not known, and is actually something one would like to infer from the data. 

Using the above observations we can qualitatively  deduce a limitation of the moment-based approach.
Roughly speaking, the information content increases with $P$, but when $P$ becomes too large the estimation breaks down. 
It exists thus a possibility that the moment error grows large before the information content of the likelihood is complete. In such situation, the moment-based likelihood cannot compete with the standard likelihood from the viewpoint of information content. If one wants to make the discussion more quantitative, one has to define the ratios
\be
\frac{I[{\cal L}_{\textrm{ mom},P}]}{I[{\cal L}_{\textrm{ std}}]}=J_P\,,\quad \frac{\hat{I}[{\cal L}_{\textrm{ mom},P}]}{I[{\cal L}_{\textrm{ mom}}]}=\hat{J}_P\,,
\ee 
where $J_P< 1$ at small $P$ and $J_P\rightarrow 1$ at large $P$,   and $\hat{J}_P\approx 1$ at small $P$ and $\hat{J}_P\neq 1$ at large $P$. $\hat{I}$ is the observed Fisher information. Some thresholds definition are then necessary to make the discussion quantitative. Here we do not go further in that direction, and focus instead on what to do in practice.

From a practical point of view,  the most robust procedure to use the moment-based likelihood seems to be as follows. 
Assume that one has a set of data at hand, and one has computed the moment-based likelihood truncated to the first $P$ moments ${\cal L}_{\rm std\,P}^{\rm  mom}$. The truncation order $P$ can be easily changed.  One then wishes to carry out a task involving the likelihood -- typically a parameter inference or a hypothesis testing, producing an output $Y$. The most robust way to proceed is to compute $Y$ for all allowed values of $P$.  That is, one starts from $P=1$, and increase $P$ until the covariance matrix becomes singular for the computing system. If a \textit{plateau} appears, this means that the information is contained in the first moments, and the value of $Y_P$ at the plateau is the one that should be kept. If no plateau appears -- because the information content does not converge fast enough before getting overridden by the error on estimation, one cannot  use reliably the moment-based likelihood.
These various behaviours will be observed in the toy-models of Sec \ref{se:toymodels}.

\subsection{Binned data}

Having derived the unbinned version of the moment-based likelihood in Eq. \eqref{eq:Lmomfull}, let us turn to the binned version. The \textit{coordinates} of the bins ${\cal D}_r$ are written as  $\bar{x}_r$. 
Estimators of the moments are then given by 
\be
\hat{m}^{\rm bin}_p= \frac{1}{n} \sum_r \hat{n}_r \bar{x}_r^p \label{eq:mom_est_bin}
\ee
where the number of events in each bin $n_r$  is normally-distributed and described by Eq. \eqref{eq:nr}. 
In these estimators, the random part is just $\hat{n}_r$,  $\bar{x}_r$ is  a fixed number. 
We have thus a linear combination of normally distributed variables. 
The $\hat{m}_p$ estimators are normally distributed and correlated to each other, such that they are described by a multivariate normal law. 
Their mean is simply 
\be
E[\hat{m}^{\rm bin}_p]= \sum_r p_r \bar{x}_r^p \,.
\ee
The covariance matrix is given by
\be
\Sigma_{pq}^{\rm bin}=\frac{1}{n}\sum_r p_r(1-p_r) \bar{x}_r^{p+q}\approx 
\frac{1}{n}\sum_r p_r \bar{x}_r^{p+q}
\,,
\ee
such that 
\be
(\hat{m}_p^{\rm bin})\sim \mathcal{N}\bigg(\sum_r p_r \bar{x}_r^p,\,\Sigma^{\rm bin}_{mn}\bigg)\,. 
\ee
An estimator of the covariance matrix is given by 
\be
\Sigma_{pq}=\frac{1}{n}(\hat{m}^{\rm bin}_{p+q}-\hat{m}^{\rm bin}_p \hat{m}^{\rm bin}_q)\,, \label{eq:cov_est}
\ee
 and the likelihood function is
\be
{\cal L}^{\rm bin}_{\rm mom}=\exp\bigg(-\frac{1}{2}(m_m^{\rm bin\, th}-\hat{m}^{\rm bin}_m)^t\, (\hat{\Sigma}^{\rm bin}_{mn})^{-1}\,(m_n^{\rm bin\, th}-\hat{m}_n^{\rm bin})\bigg)\,.
\label{eq:Lmombin}
\ee
It has the same structure as in the unbinned case.

The information content of this likelihood is somewhat simpler to understand than the one for the unbinned moment-based likelihood. 
Assume data are binned with $R$ the number of bins. Then the number of moments cannot exceed $R$. Otherwise, any extra moment can be written as a linear combination  of the previous ones, such that the moment covariance matrix becomes singular. 
This is also reminiscent from a version of the Nyquist-Shannon's sampling theorem applied to a discrete Laplace transform. For $n$ sampled points of the distribution, exactly $n$ moments are sufficient to fully reproduce the distribution.
We have checked this behaviour on binned toy-models.
We do not explore further this direction in this paper, focusing instead on the unbinned likelihoods.

\subsection{The multivariate case}

So far we considered the shape analysis of a univariate distribution of data. Our approach readily generalizes to an arbitrary number of observables $D$.  The vector of moments is replaced by a rank-$D$ tensor, and the moment covariance matrix is replaced by a rank-$2D$ tensor. Labelling the $D$ different observables as $X_{(D)}$,  the joint moment estimators are 
\be
\hat{m}_{p_1\ldots p_D}= \frac{1}{n}\sum_{i=1}^{n} X_{(1)\,i}^{p_1}\ldots X_{(D)\,i}^{p_D}
\ee
The covariance tensor is 
\be
\Sigma_{p_1\ldots p_D,q_1\ldots q_D}= \hat{m}_{p_1+q_1\ldots p_D+q_D}-\hat{m}_{p_1\ldots p_D}\hat{m}_{q_1\ldots q_D}\,.
\ee
For example for the 2D case,  $\Sigma_{pp',qq'}=\hat{m}_{p+q,p'+q'}-\hat{m}_{p,p'}\hat{m}_{q,q'}$.
The covariance tensor is symmetric under the exchange of the two blocks of indexes, 
\be
\Sigma_{p_1\ldots p_D,q_1\ldots q_D}=\Sigma_{q_1\ldots q_D,p_1\ldots p_D}\,.
\ee
Using characteristic functions, the Central Limit Theorem applies similarly to the 1D case, except that one needs to define carefully the generalized inverse of the covariance tensor. We find
\be
{\cal L}_{\rm mom }=\exp\bigg(  -\frac{1}{2} ( \hat{m}_{p_1\ldots p_D}-m_{p_1\ldots p_D} ) \Sigma^{-1}_{p_1\ldots p_D,q_1\ldots q_D}  ( \hat{m}_{q_1\ldots q_D}-m_{q_1\ldots q_D} )    \bigg)\,,
\ee
where the inverse covariance tensor satisfies
\be\Sigma^{-1}_{p_1\ldots p_D,x_1\ldots x_D} \Sigma_{x_1\ldots x_D,q_1\ldots q_D}=\delta_{p_1 q_1}\ldots \delta_{p_D q_D}\,. \ee
The only technical complication with respect to the 1D case is the computation of the inverse covariance tensor. Some   machinery may be required to carry out this task efficiently. 
We  focus on the 1D case for the rest of the paper.

\section{Shape analysis for signal searches}  \label{se:signal}

Having laid out the general features of standard and moment-based likelihoods of shape analysis, let us focus on the typical scenario of high-energy physics. 
What happens typically in higher-energy data analysis is the search for a small signal over a background. 
If the new effect researched is the decay of a somewhat stable new particle, the signal has the form of  a narrow Lorentzian, and appears on the top of a broader background. This is for example how the Higgs has been found at the LHC. Apart from this particular case where the signal is a ``bump'', a new physics signal  can take in general an arbitrary form.

Given that no light new physics beyond the SM has been found so far at the LHC, the scenario of a heavy new physics is fairly preferred by current observations. Whenever the mass scale of the new physics effect is higher than the experiment energy, the low-energy effects of new physics  can be enclosed into effective operators of higher dimension  that supplement the SM Lagrangian. They are suppressed by powers of the new physics scale $\Lambda$, for example dimension six operators have the form $\alpha/\Lambda^2 \,\mathcal{O}$ (see e.g. \cite{Buchmuller:1985jz,Grzadkowski:2010es} for the complete SM basis, \cite{Masso:2014xra} for a recent review).
 Searching for these operators and inferring knowledge about both $\alpha$ (see e.g.\cite{Dumont:2013wma,Ellis:2014dva}
) and $\Lambda$ \cite{NP_scale} can be considered 
as a major line for current and future new physics searches. These effective operators contribute to create or modify the matrix elements that describe particle reactions,
$\mathcal{M}_{\rm SM}+\alpha/\Lambda^2\mathcal{M}_{\rm NP}$ (see e.g. \cite{Chen:2014xra} for double Higgs production). Notice that $\mathcal{M}_{\rm NP}$ may or not interfere with $\mathcal{M}_{\rm SM}$. No resonance can be produced in such scenario. Instead, the effective operators typically induce broad deviations, that need to be detected over a broad background. Shape analysis has therefore an important role to play in this precision physics program.
Although the likelihoods and results we present below are slightly oriented toward high-energy signal searches, they can be used independently of the physical context. All the results presented below follow a general parametrization, independent of the physics.

Let us  consider that the data available are distributed over a variable $x$ in a domain $\cal D$, following a (un-normalized) distribution denoted $\hat{d}$.~\footnote{Note for unbinned data, $\hat d$ can be represented as a sum of Dirac delta associated to each event,
$\hat d= \sum_i \delta(X_i-x)$. }
The hypothetical distribution one wants to compare to the data can be written as
\be
d=d_{b}+\mu\, d_s\,,
\ee
where $\mu$ is the signal strength. This is the parameter of interest we want to gain knowledge about.
 $d_{b}$ is the expected background, and  $d_s$ is the signal predicted by the hypothesis. Setting $\mu=0$ corresponds to testing the background-only hypothesis. Setting $\mu=1$ corresponds to testing the predicted value of the signal.
These definitions match the usual formalism for global event rates. The event rates are obtained by summing all events over $\cal D$,
\be
\hat{n}=\int_{\cal D} \hat{d}\,dx \,\quad n=\int_{\cal D} d\, dx
\,\quad n_b=\int_{\cal D} d_b\, dx\,\quad n_s=\int_{\cal D} d_s\, dx\,,
\ee
which gives the usual, familiar parametrization for signal searches
\be
n=n_{b}+\mu\,n_s\,.
\ee

Let us now go beyond the global event rates, and analyse the shape of the data along $x$. 
Again we focus on the case of normalized distributions, which do not include the total event rates. The formulas including the total event rate are obtained very similarly, and discussed in the last subsection.
 Both observed and hypothetical distributions have to be normalized to one, and one defines the PDFs for background and signal,
\be
\hat{f}=\frac{\hat{d}}{\hat{n}}\,, \quad f=\frac{d}{n}\,, \quad f_b=\frac{d_b}{n_b}\,, \quad f_s=\frac{d_s}{n_s}\,.
\ee
The hypothetical data shape takes therefore the form
\be
f=\frac{n_b\,f_b+\mu \,n_s\, f_s}{n_b+\mu\, n_s}\,.
\ee
This is the central quantity for signal searches through shape analysis.
Note in cases where the expected signal rate is small with respect to the background,  $\mu n_s \ll n_b$,  , which is the typical situation for signal searches, the shape takes the form
\be
f=f_b+\mu \,\frac{n_s}{n_b}\,( f_s -f_b)+ O\bigg(\frac{\mu^2 n_s^2}{n_b^2} \bigg)\,.
\ee
We can now build the various likelihoods introduced in Secs.~\ref{se:like_std}, \ref{se:like_mom}. We omit the hat over $\hat n$ from now on.
We also display the signal strength given by the maximum likelihood (ML) estimator $\hat \mu$ and its associated variance $\hat \sigma^2$. When the data sample is large enough, these can be directly used in simple likelihood-based statistical tests. 
With a large enough sample, one can expand around $\hat \mu$ (see Wilk-Wald's theorems \cite{Wilk,Wald}) and $(\mu-\hat\mu)^2/\hat \sigma^2$ follows a chi-squared law with one degree of freedom. 
Defining the significance of the statistical test as $Z=\Phi^{-1}(1-p)$, where $\Phi$ is 1D the cumulative normal distribution with standard deviation and $p$ the p-value of the test, one simply has (see \cite{asimov} for an enlightening discussion of all the possibilities)
\be
Z_\mu=\Phi^{-1}\bigg(2\Phi\bigg(\frac{|\hat{\mu}-\mu|}{\hat\sigma}\bigg)-1\bigg)\,,\quad Z=\frac{\hat{\mu}}{\hat\sigma}\,\quad  \label{eq:Z}\,,
\ee
respectively for  $\mu$ with both signs allowed, and for the discovery of a positive signal.


\subsection{Standard likelihood}

The standard unbinned likelihood reads 
\be
{\cal L}_{\rm std}=\prod_i^n \frac{n_b\,f_b(X_i)+\mu \,n_s\, f_s(X_i)}{n_b+\mu\, n_s}\,.
\ee
From it we can infer information on the signal strength.
 The signal strength at the maximum likelihood  $\hat \mu$ cannot be put under a close form in the general case. However it is worth noticing that when the condition $\mu f_s \ll f_b$ holds for any measured $X_i$ (this condition is much stronger than $\mu\,n_s\ll n_b$), the maximum likelihood signal strength takes the form 
\be
\frac{\hat{\mu}}{\hat{\sigma}^2}=\frac{n_s}{n_b}\sum_i \left(\frac{f_s(X_i)}{f_b(X_i)}-1\right)\,,\quad \frac{1}{\hat{\sigma}^2}=\frac{n_s^2}{n_b^2}\sum_i \left(\frac{f_s(X_i)}{f_b(X_i)}-1\right)^2\,.
\ee


Let us turn to binned data. One defines the observed and expected event probabilities over each bin $\mathcal{D}_r$,
\be
\hat{p}_r= \int _{{\cal D}_r} dx\, \hat{f}\,\quad p_{s,r}=  \int _{{\cal D}_r} dx\, f_s\,\quad p_{b,r}=  \int _{{\cal D}_r} dx\, f_b\,.
\ee

The standard binned likelihood reads
\be
{\cal L}_{\rm std}^{\rm bin}=\prod_i \exp\bigg(-   \bigg( \frac{n_b p_{b,r}+\mu\,n_s p_{s,r}  }{n_b+\mu\,n_s} -\hat{p}_r\bigg)^2\, \frac{n}{2\,\hat{p}_r}  \,\bigg)\,.
\ee
For $\mu\,n_s \ll n_b$, it simplifies to 
\be
{\cal L}_{\rm std}^{\rm bin}=\prod_i \exp\bigg(-  \bigg( \mu\, \frac{n_s}{n_b} ( p_{s,r}- p_{b,r})   -\Delta\hat{p}_r\bigg)^2\, \frac{n}{2\,\hat{p}_r} \,\bigg)\,,
\ee
where one defined the observed deviation $\Delta\hat{p}_r=\hat{p}_r-p_{b,r} $ (or $\Delta\hat{n}_r=\hat{n}_r-n_{b,r} $).

The ML signal strength  $\hat{\mu}$ and the variance $\hat\sigma^2_{\rm bin}$ read
\be
\hat{\mu}_{\rm bin}= \frac{n_b}{n_s}\, \sum_r \frac{(p_{s,r}-p_{b,r})\Delta_{\hat{p}_r}}{\hat{p}_r} \bigg[
\sum_r \frac{(p_{s,r}-p_{b,r})^2}{\hat{p}_r}\bigg]^{-1}\,,\quad \frac{1}{\hat\sigma^2_{\rm bin}}= n\,\frac{n_s^2}{n_b^2}\,
\sum_r \frac{(p_{s,r}-p_{b,r})^2}{\hat{p}_r}
\ee
 They will appear in the statistical tests.  Note the $n_s/n_b$ factors will always cancel for the discovery test of Eq.~\eqref{eq:Z}.

\subsection{Moment-based likelihood}

The moments of the data distribution are given by Eq.\eqref{eq:mom_est_unb}. The moments of the hypothetical shape are expressed in terms of the background and signal moments as
\be
m_p=\frac{n_b m_{b,p}+\mu\,n_s m_{s,p}}{n_b+\mu\,n_s}\,.
\ee
The exact moment-based likelihood for unbinned data reads therefore
\be
{\cal L}_{\rm mom}=\exp\bigg(- \frac{n}{2} \bigg( \frac{n_b m_{b,p}+\mu\,n_s m_{s,p}}{n_b+\mu\,n_s}-\hat{m}_p  \bigg) \bigg[ \hat{m}_{p+q}-\hat{m}_p \hat{m}_q \bigg]^{-1}_{pq}   \bigg(  \frac{n_b m_{b,q}+\mu\,n_s m_{s,q}}{n_b+\mu\,n_s}-\hat{m}_q  \bigg) \bigg)
\ee
For a small signal $\mu\,n_s \ll n_b$, one defines the observed deviations $\Delta\hat{m}_p=\hat{m}_p-m_{b,p} $, and the likelihood simplifies to 
\be
{\cal L}_{\rm mom}=\exp\bigg(- \frac{n}{2} \bigg( \mu\,\frac{n_s}{n_b} (m_{s,p}-m_{b,p} )-\Delta\hat{m}_p  \bigg) \bigg[ \hat{m}_{p+q}-\hat{m}_p \hat{m}_q \bigg]^{-1}_{pq}   \bigg( \mu\,\frac{n_s}{n_b} (m_{s,q}-m_{b,q})-\Delta\hat{m}_q  \bigg) \bigg)\,.
\ee
The ML signal strength and the associated variance appear to be
\be
\frac{\hat{\mu}_{\rm mom}}{\hat\sigma^2_{\rm mom}}=  n \,\frac{n_s}{n_b}\,
(m_{s,p}-m_{b,p}) \bigg[ \hat{m}_{p+q}-\hat{m}_p \hat{m}_q \bigg]^{-1}_{pq}  \Delta\hat{m}_{q}\,.
\ee
\be
\frac{1}{\hat\sigma^2_{\rm mom}}= n\,\frac{n_s^2}{n_b^2}\, (m_{s,p}-m_{b,p}) \bigg[ \hat{m}_{p+q}-\hat{m}_p \hat{m}_q \bigg]^{-1}_{pq} ( m_{s,q}-m_{b,q})\,,\ee

Let us turn to binned data. The  moment estimators are given by Eq.~\eqref{eq:mom_est_bin}. The moments of the hypothetical distribution are given by 
\be
m_p^{\rm bin}=\sum_r \frac{n_b\, p_{b,r}+\mu\, n_s\, p_{s,r}}{n_b+\mu\, n_s} \bar{x}_r^p=\sum_r \bigg( p_{b,r}+\mu\, \frac{n_s}{n_b}\, (p_{s,r}-p_{b,r})+O\bigg(\frac{\mu^2 n_s^2}{n_b^2}\bigg) \bigg)\bar{x}^p_r\,.
\ee
Introducing the observed deviations $\Delta \hat{n}_r= \hat{n}_r-n_{b,r}$, the likelihood for small signal reads 
\be
{\cal L}_{\rm mom}^{\rm bin}=\exp\bigg(- \frac{n}{2} \bigg( \mu \frac{n_s}{n_b} ( m_{s,m}-m_{b,m} )-\Delta\hat{m}_m  \bigg) \bigg[ \hat{m}_{m+n}-\hat{m}_m \hat{m}_n \bigg]^{-1}_{mn}   \bigg(  \mu \frac{n_s}{n_b} ( m_{s,n}-m_{b,n} )-\Delta\hat{m}_n  \bigg) \bigg)\,.
\ee
The ML signal strength and the variance are
\be
\frac{\hat{\mu}^{\rm bin}_{\rm mom}}{(\hat\sigma^{\rm bin}_{\rm mom})^2}=  n\,\frac{n_s}{n_b}\,
(m_{s,p}^{\rm bin}-m_{b,p}^{\rm bin}) \bigg[ \hat{m}_{p+q}^{\rm bin}-\hat{m}_p^{\rm bin} \hat{m}_q^{\rm bin} \bigg]^{-1}_{pq}  \Delta\hat{m}_{q}^{\rm bin}\,. \label{eq:mumom}
\ee
\be
\frac{1}{(\hat\sigma^{\rm bin}_{\rm mom})^2}= n\,\frac{n_s^2}{n_b^2}\, (m_{s,p}^{\rm bin}-m_{b,p}^{\rm bin}) \bigg[ \hat{m}_{p+q}^{\rm bin}-\hat{m}_p^{\rm bin} \hat{m}_q^{\rm bin} \bigg]^{-1}_{pq} ( m_{s,q}^{\rm bin}-m_{b,q}^{\rm bin})\,.
\label{eq:sigmamom}
\ee

\subsection{Combining shape and event rate }

In this work we write explicitly the  likelihoods for shape-information only. In general one may also want to include the event rates in an analysis. 
For the standard likelihoods, given the Poisson nature of the data, the likelihood with both shape and event rate  reads 
\be
\mathcal{L}=\mathcal{L}_{\rm std}\, \mathcal{L}_{\rm tot}\,,
\ee
where
$\mathcal{L}_{\rm std}$ is the shape-only likelihood defined in Eq.~\eqref{eq:Lfull} and \be
\mathcal{L}_{\rm tot}=(n_b+\mu\,n_s)^{\hat n}\, e^{ -( n_b+\mu\, n_s )}\,.
\ee
This combination is exact. 
For the unbinned moment-based likelihood $\mathcal{L}_{\rm mom}$, whenever it is a good approximation of $\mathcal{L}_{\rm std}$, it can be combined with $\cal{L}_{\rm tot}$ in the same way. Another way to include the event rate in $\mathcal{L}_{\rm mom}$  is to work with the un-normalized  observed and hypothetical distributions $\hat{d}$ and $d$. 
For the binned likelihoods, $\cal{L}_{\rm std}^{\rm bin}$, $\cal{L}_{\rm mom}^{\rm bin}$, the most direct way to include the global event rate is also to use un-normalized distributions. For  the small signal results obtained in the subsections above, this amounts to do the replacement $n_{s,r}-n_{b,r}\rightarrow n_{s,r}$, $m_{s,p}-m_{b,p}\rightarrow m_{s,p}$.
 One can also  include the information about the event rate by multiplying $\mathcal{L}^{\rm bin}$ by $\mathcal{L}_{\rm tot}$. These different approaches are not formally equivalent
 and may let appear small discrepancies, unless either the event rate or the shape information dominate the information content.

\section{Advantages and limits of  the moment-based likelihood \label{se:limits}}

The information content of the standard likelihood cannot be improved in any alternative approach. Rather, the main advantage of the moment-based likelihood resides at the technical level, as it can simplify the  process of shape analysis. Before  discussing further this practical aspect, let us understand in which situation the moment-based likelihood can compete with the standard likelihood. 

For a given background, there is in principle an infinity of signal shapes possible. Without specifying any detail of the shapes, one can roughly classify the signals depending whether it is localized over the background, or if instead it appears as a broad, overall deformation of the background.
   Let us denote by $\mathcal{D}$ the support of the data distribution, and denote the restriction of a distribution $g$ to a domain $\mathcal{D}'$ as $g_{\mathcal{D}'}$.
\paragraph*{Definition\,\,1} If it exists a subdomain $\mathcal{D}'\subset\mathcal{D}$ such that
$(f_s/f_b)_{\mathcal{D'}}\gg (f_s/f_b)_{\mathcal{D\backslash D'}}$, 
 the signal is said to be \textit{local} in $\mathcal{D}$. If no subdomain $\mathcal{D}'\subset\mathcal{D}$ exists such that  $(f_s/f_b)_{\mathcal{D'}}\gg (f_s/f_b)_{\mathcal{D\backslash D'}}$,  
 the signal is then said to be \textit{non-local} in $\mathcal{D}$.

Such classification is only qualitative, and could certainly be refined. However it is sufficient for our purposes. 
We use it to make the following qualitative argument. We have seen in Sec.~\ref{se:like_mom} that these are the first moments of the moment sequence which are the best estimated, and which enter in the likelihood in practice. 
 By definition, the first moments characterize the global, \ie~non-local features of a distribution. Therefore the moment-based likelihood should contain as much information as the standard likelihood for non-local signals. Instead, for local signals, one expects the performance of the moment-based likelihood to decrease with respect to the ones of the standard likelihood.
Examples of local signals are ``bumps'' and ``fat tails'', that will appear in the toy-models of Sec. \ref{se:toymodels}

Let us remark that we did not derive ${\cal L}_{\rm mom}$ directly from ${\cal L}_{\rm std}$ in Sec.~\ref{se:like_mom}.
Such a derivation does not seem to be straightforward. If it exists, it may help defining more precisely the condition for having ${\cal L}_{\rm mom}$ (approximately) equivalent to ${\cal L}_{\rm std}$. 
For the present work we do not go further in that direction and leave this derivation as an interesting open problem. 
A related issue is the behaviour of the significance in the ``fat tail'' case, see Sec.~\ref{se:toymodels}.

Let us now discuss in details the practical interest of the moment-based likelihood. 
One of the advantages of the moment-based likelihood is purely technical. 
Quite often, the exact analytical form of the hypothetical distribution of background and signal $f_b$ and $f_s$ is \textit{unknown}. Rather, they need to be evaluated using Monte-Carlo simulations. Once these simulations of pseudo-data are done, the task remains  of  obtaining some analytical expression of $f_b$ and $f_s$, that one needs to plug in $\cal L_{\rm std}$ . The most simple and common technique seems to be the use of binning. But the problem with such straightforward method is that it always induces  a loss of  information, as discussed in Sec \ref{se:like_std}. 
To estimate analytical expressions for $f_b$ and $f_s$ without information loss, one has therefore to rely on more evolved techniques of fitting, like kernel density estimation. 
However this step of fitting remains tricky, whatever the technique, and needs careful cross-validation. Indeed, any small error of the fit of the background can potentially spoil the search for the signal. That is, as both fitting errors and signal potentially look like a small deformation of $f_b$,  a slight error in the fit can be misinterpreted as a signal.
Notice that in general the problem of fit errors drastically increases with the number of dimensions.

Remarkably,  the moment-based likelihood  bypasses this tricky step of fitting the expected $f_b$ and $f_s$ shapes. Indeed, once the densities are obtained from Monte-Carlo simulations, it is straightforward to deduce the moments of the expected $f_b$ and $f_s$. The uncertainty associated with the MC estimation of the moments is described by a covariance matrix like Eq.~\eqref{eq:cov_est}, suppressed by the total number of events of the simulation $n_{MC}$. To make sure that the MC uncertainty is well negligible with respect to the actual statistical uncertainty, the criteria is simply that $n_{MC}\gg n$. The MC uncertainty is thus easily kept under control. Finally, notice that the  step of precisely fitting the MC results gets increasingly trickier in higher dimensions, and slight fitting errors are more likely to happen.
 The moment-based approach becomes thus even more attractive in that case. 

Depending on the scenario of search,  one may or may not know in advance whether the signal is  local in the sense of Def.~$\textcolor{blue}{1}$. If one knows that the signal is non-local, the standard likelihood can be just replaced with the moment-based likelihood, with the technical benefits described above. On the other hand, if one knows that the signal is local, the standard likelihood is expected to give better results. 

An interesting possibility appears if one does not know in advance whether the signal is  local or not. Let us assume that one is performing a discovery test, aiming at excluding the background-only hypothesis. 
No assumption is made on the form of the signal. Let us now assume that both $\mathcal{L}_{\rm mom}$ and $\mathcal{L}_{\rm std}$ are computed (without fitting error for the latter). The significances of the discovery tests are denoted $Z_{\rm mom}$, $Z_{\rm std}$, and the standard significance $Z_{\rm std}$ is assumed to point toward the existence of a signal. Then, getting $Z_{\rm mom}\approx Z_{\rm std}$ implies that the signal is non-local while getting $Z_{\rm mom}\neq Z_{\rm std}$ implies that the signal is local. That is, one gets a useful information on the shape of the signal, using only a discovery test with two different likelihoods. 

Imagine for example that an effective operator $\mathcal{L}_{\rm eff}\supset \alpha/\Lambda^2\, \mathcal{O}$ is expected to modify the shape of the signal. It interferes with the background, such that for a given sign of $\alpha$ the signal is local (e.g. a fat tail), while for the other sign of $\alpha$ the signal is non-local. Then the test we described in the paragraph above readily provides a discrimination on the sign of the effective operator. The knowledge of the sign of an effective operator  can translate as a powerful constraint on the models that contribute to $\mathcal{O}$.

\section{Toy-models for signal searches \label{se:toymodels}}

\begin{figure}
\begin{picture}(400,150)
\put(0,0){		\includegraphics[trim=0cm 0cm 0cm 0cm, clip=true,width=4.9cm]
		{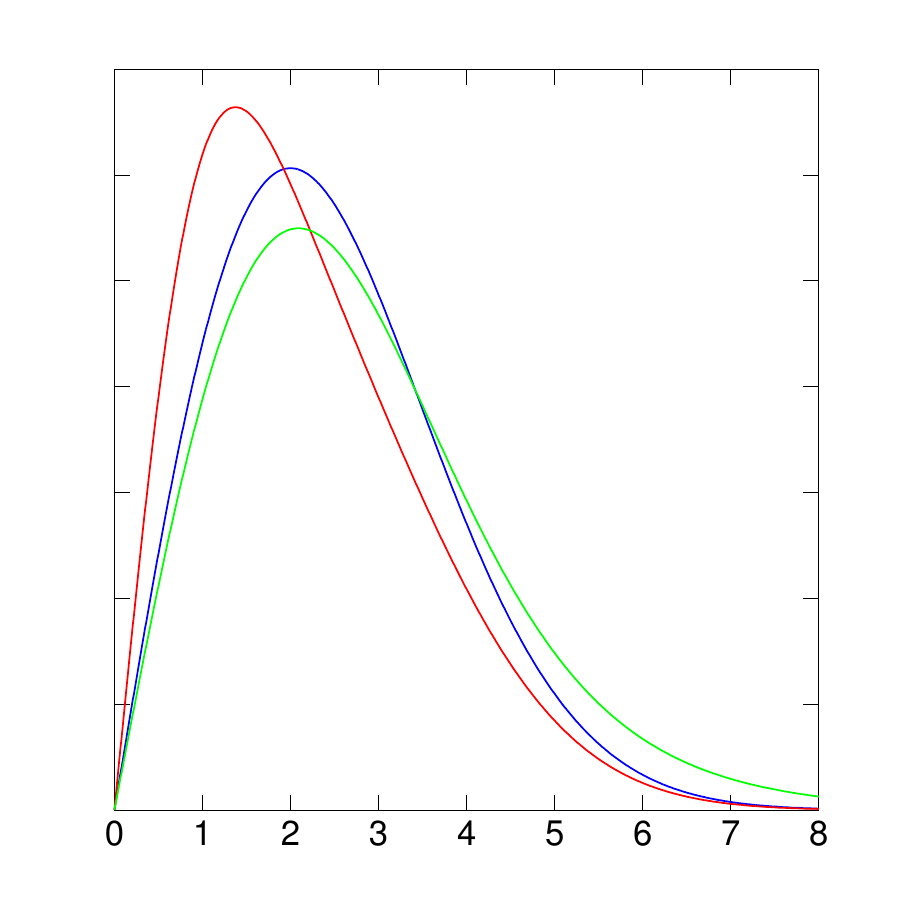}		}
\put(140,0){				\includegraphics[trim=0cm 0cm 0cm 0cm, clip=true,width=4.9cm]
		{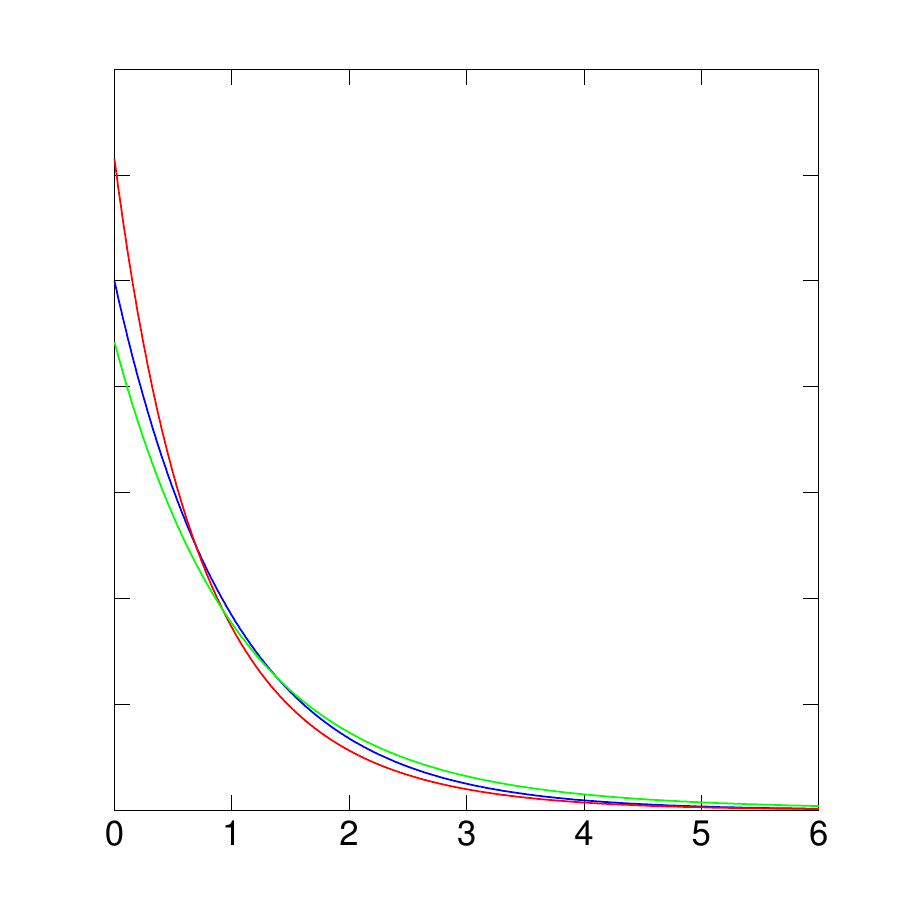}		}
\put(280,0){			\includegraphics[trim=0cm 0cm 0cm 0cm, clip=true,width=4.9cm]
		{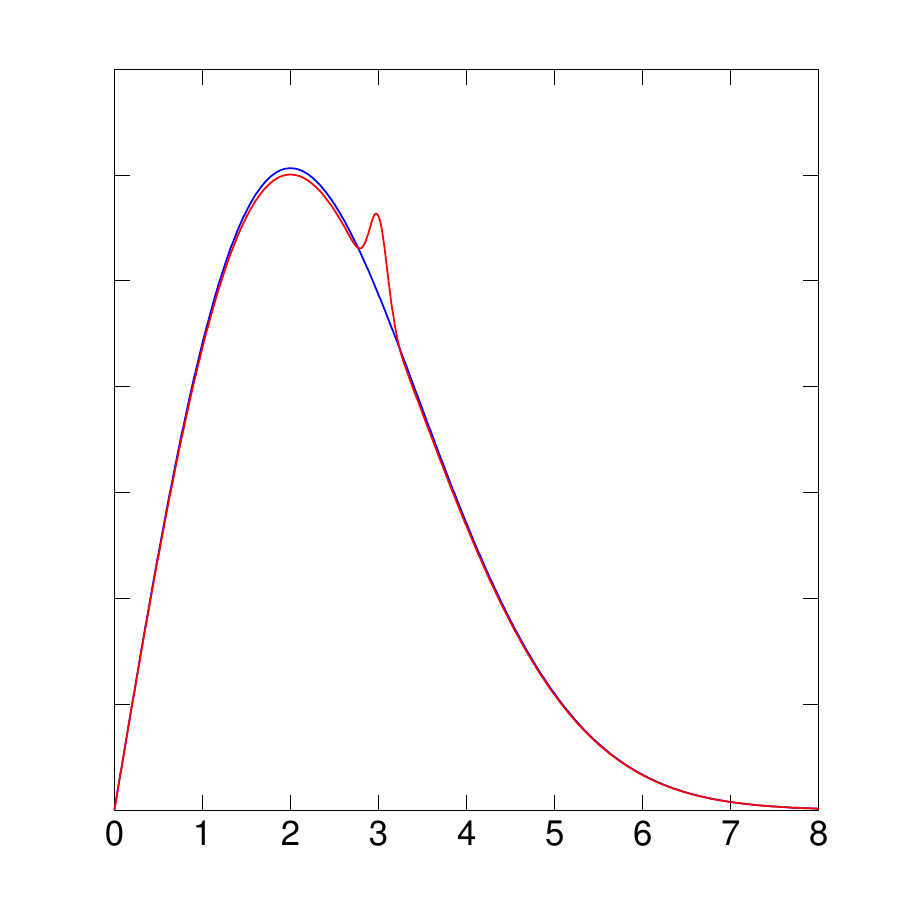}	    }
\put(70,0){$x$}		\put(210,0){$x$}			\put(350,0){$x$}		
		\end{picture}
		\vspace{0.0cm}
	\caption{ Examples of toy-models for signal searches. \textit{Left}: Rayleigh signals with $\rho=1$ (red), $\rho=3$  (green), with  strength $\mu=30\%$ over a Rayleigh background with $\rho=2$ (blue).
	\textit{Center}: exponential signals with $\lambda=2$ (red), $\lambda=0.5$ (green), with  strength $\mu=30\%$ over an exponential background with $\lambda=1$ (blue).
	\textit{Right}: A bump.
		 \label{fig:shaperef}}
\end{figure}
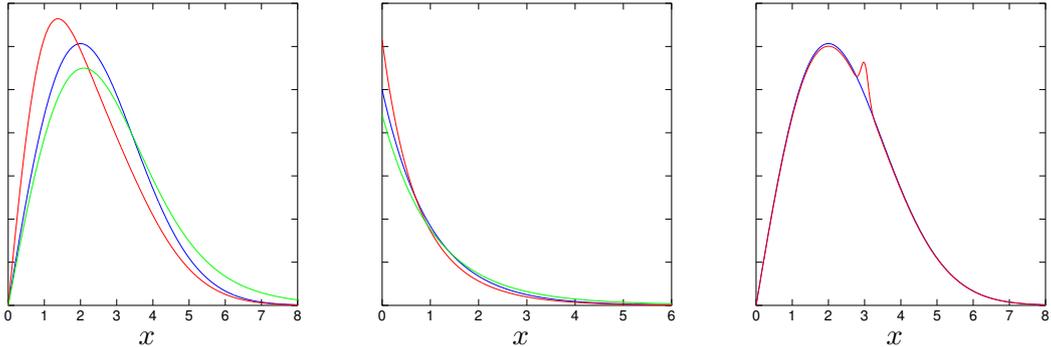

In this Section we perform signal searches within various toy-models, using both standard and moment-based techniques. This serves to both check and exemplify the formalism and methods introduced in Secs.~\ref{se:like_std} to  \ref{se:limits}.
The toy-models are chosen in order to mimick typical distributions obtained from  LHC measurements.
 We focus on the search for a signal in 1D data distributions. The observable is denoted $X$, and the pseudo-data PDF is denoted $\hat{f}_X(x)$, consistently with Sec.~\ref{se:signal} notations. The distributions and parameters used to generate the pseudo-data will be denoted by a \textit{tilde} (these are not observed quantities, so they should not be hatted).  
The amount of background and signal events introduced in the data sample are written as $\tilde{n}_b$, $\tilde{n}_s$. The hypothesized event numbers $n_b$, $n_s$ will not appear below because they vanish in the discovery test we are going to use. 
  The pseudo-data are generated using the following toy-models:
\begin{itemize}
\item A Rayleigh background with a Rayleigh signal, with respective shape parameters $\tilde{\rho}_b$, $\tilde{\rho}_s$\,,
\item An exponential background with an exponential signal, with respective shape parameters $\tilde{\lambda}_b$, $\tilde{\lambda}_s$\,,
\item A Rayleigh background with a Gaussian bump\,.
\end{itemize}
These various configurations are displayed in Fig. \ref{fig:shaperef}.
Formulas for the various PDFs and moments are collected in App. \ref{app:distributions}.
In what follows, the shape parameters for data and hypothetical distributions will always be the same, so that we will drop their tilde from now on. 

From the point of view of Def. \textcolor{blue}{1}, one can roughly say that the signal is local when the background and signal shape parameters are not too different. If the Rayleigh (resp. exponential) data  have $\rho_s\ll\rho_b$ (resp. $\lambda_s\gg\lambda_b$), the signal is peaked over the background, so it is local. If $\rho_s\gg\rho_b$ and $\lambda_s\ll\lambda_b$, the tail of the signal at large $X$ is large with respect to the tail of the background, so again the signal is local. We denote this case as a ``fat tail'' signal. 
Notice in our toy-models one has actually $f_s/f_b\rightarrow \infty$ for large $x$ in this regime.

We generate a larger number of pseudo-data for the background and signal, compute the $\rm p$-value  and the equivalent significance for a test of the \textit{discovery of the signal}, as described in Sec.~\ref{se:signal}.
We use the discovery test of Eq.~\eqref{eq:Z} together with the ML estimators of the unbinned moment-based likelihood Eq.~\eqref{eq:mumom},\eqref{eq:sigmamom},
\be
Z_{\rm mom}=\frac{\hat \mu_{\rm mom}}{\hat\sigma_{\rm mom}}\,.
\ee
For the purpose of testing the moment-based likelihood, we also compute for each pseudo-experiment the significance $Z_{\rm std}$ given by the standard likelihood. For that purpose,  $\hat{\mu}$ and $\hat{\sigma}$ are obtained by maximizing the negative log-likelihood and taking the second derivative,
\be
\frac{\partial}{\partial \mu}\log\mathcal{L_{\rm std}}|_{\mu=\hat\mu_{\rm std}}=0\,,\quad -\frac{\partial^2}{\partial \mu^2}\log\mathcal{L}\bigg|_{\mu=\hat\mu_{\rm std}}=\frac{1}{\hat{\sigma_{\rm std}}^2}\,,\quad
Z_{\rm std}=\frac{\hat \mu_{\rm std}}{\hat\sigma_{\rm std}}\,.
\ee

%


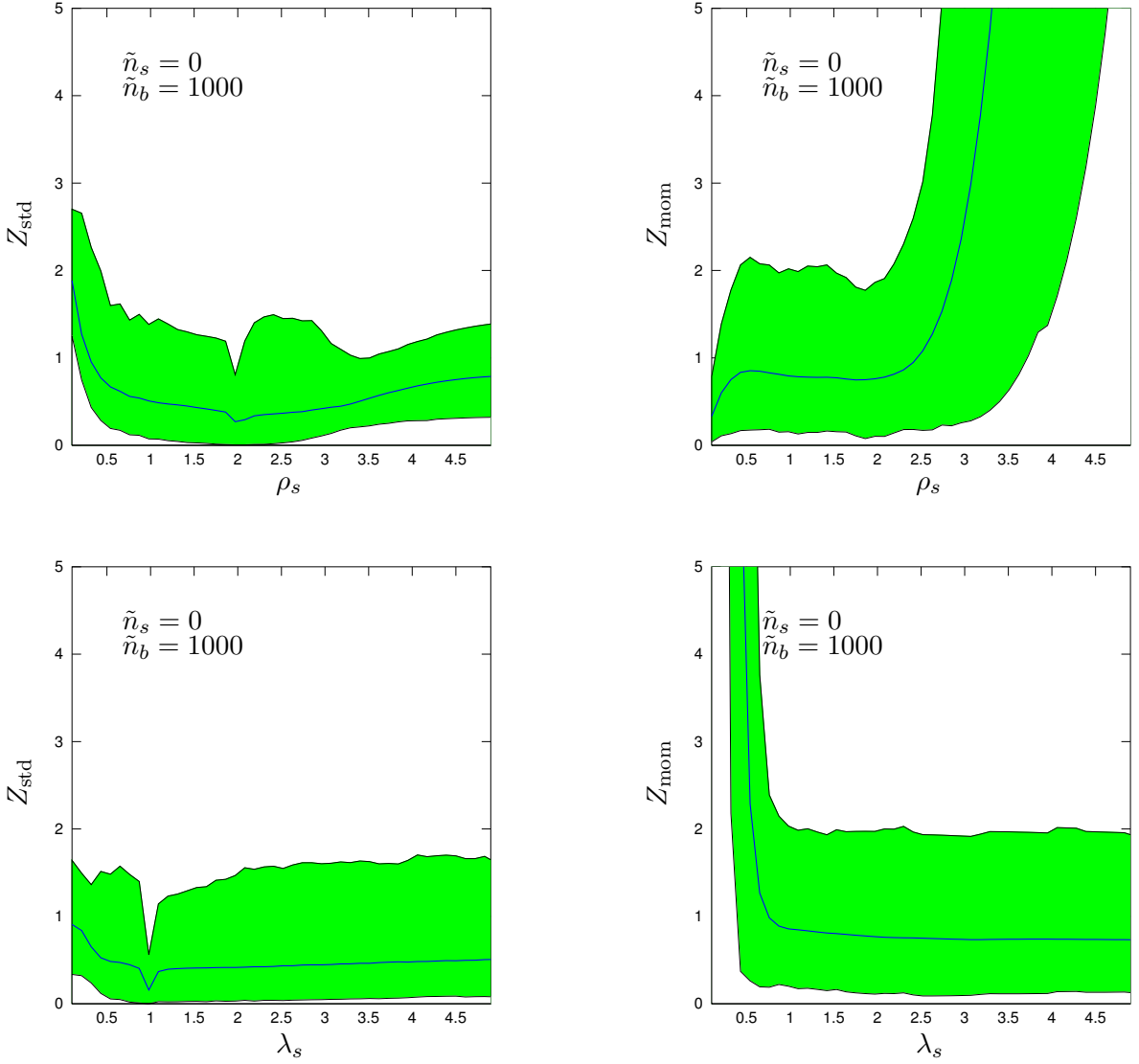
\begin{figure}
\begin{picture}(400,400)
\put(0,220){			\includegraphics[trim=0cm 0cm 0cm 0cm, clip=true,width=7.4cm]
		{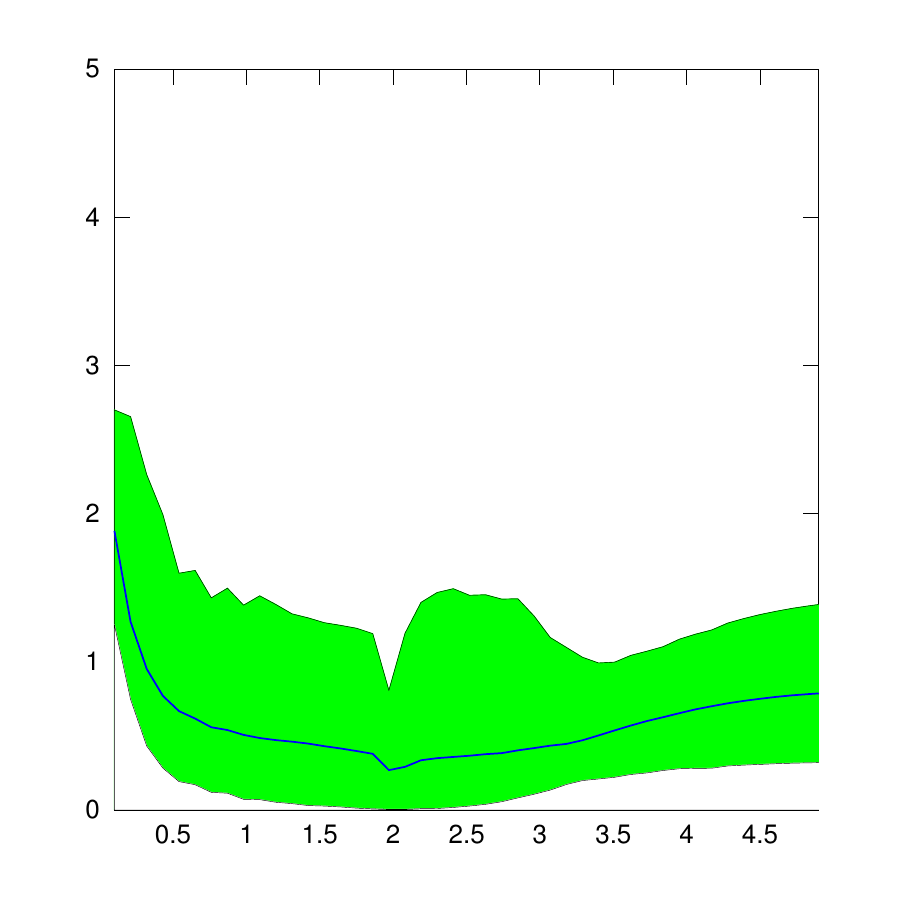}	}
							\put(50,390){$\tilde{n}_s=0$}	\put(50,380){$\tilde{n}_b=1000$}
							\put(110,225){$\rho_s$} \put(5,320){\rotatebox{90}{$Z_{\rm std}$}}
	\put(250,220){							\includegraphics[trim=0cm 0cm 0cm 0cm, clip=true,width=7.4cm]
		{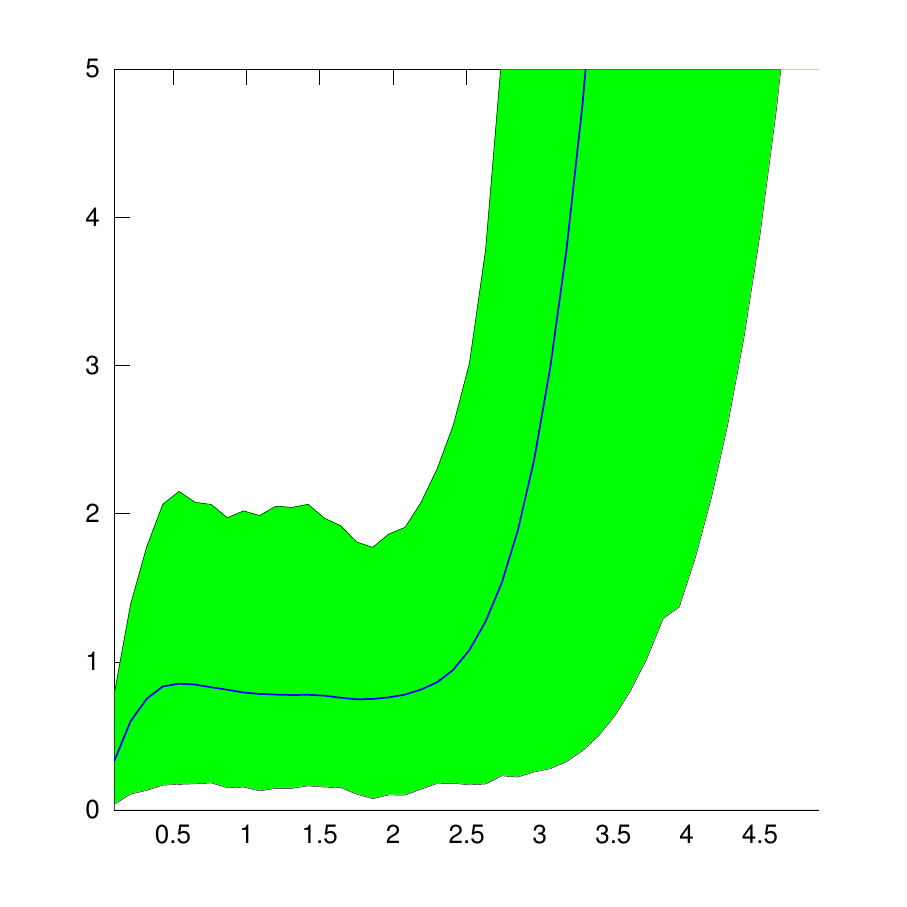}	 }
								\put(300,390){$\tilde{n}_s=0$}	\put(300,380){$\tilde{n}_b=1000$}
								\put(360,225){$\rho_s$} \put(255,320){\rotatebox{90}{$Z_{\rm mom}$}}
\put(0,0){			\includegraphics[trim=0cm 0cm 0cm 0cm, clip=true,width=7.4cm]
		{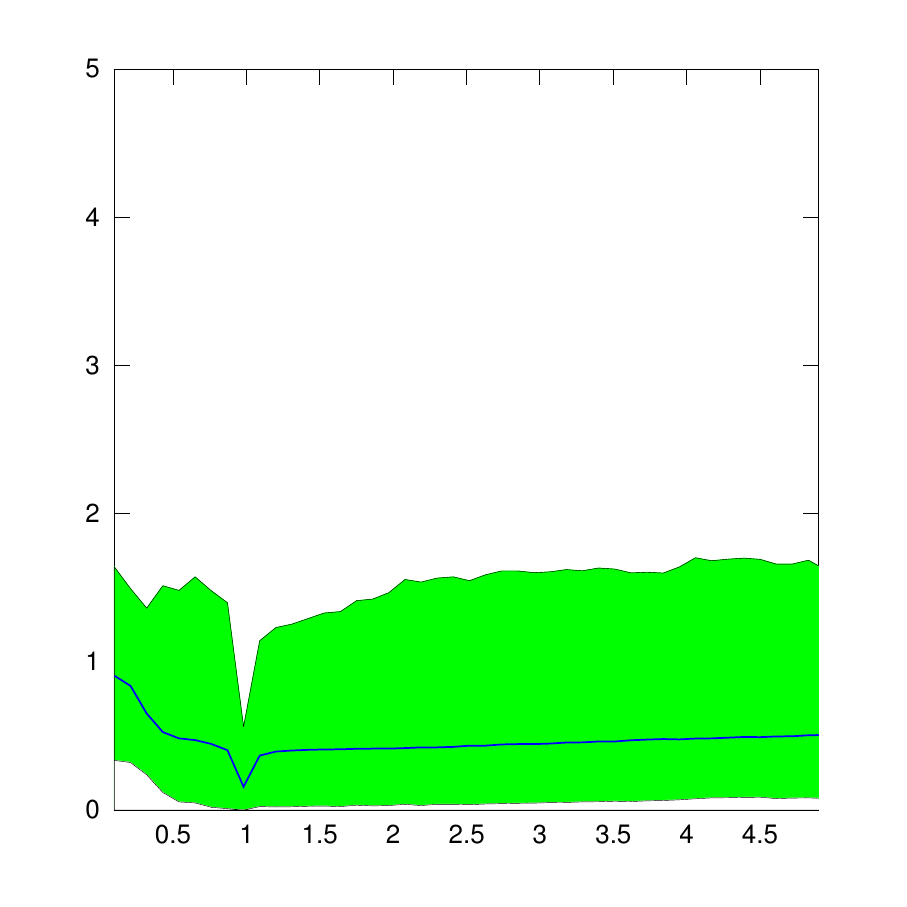}	}
							\put(50,170){$\tilde{n}_s=0$}	\put(50,160){$\tilde{n}_b=1000$}
							\put(110,3){$\lambda_s$} \put(5,100){\rotatebox{90}{$Z_{\rm std}$}}
	\put(250,0){							\includegraphics[trim=0cm 0cm 0cm 0cm, clip=true,width=7.4cm]
		{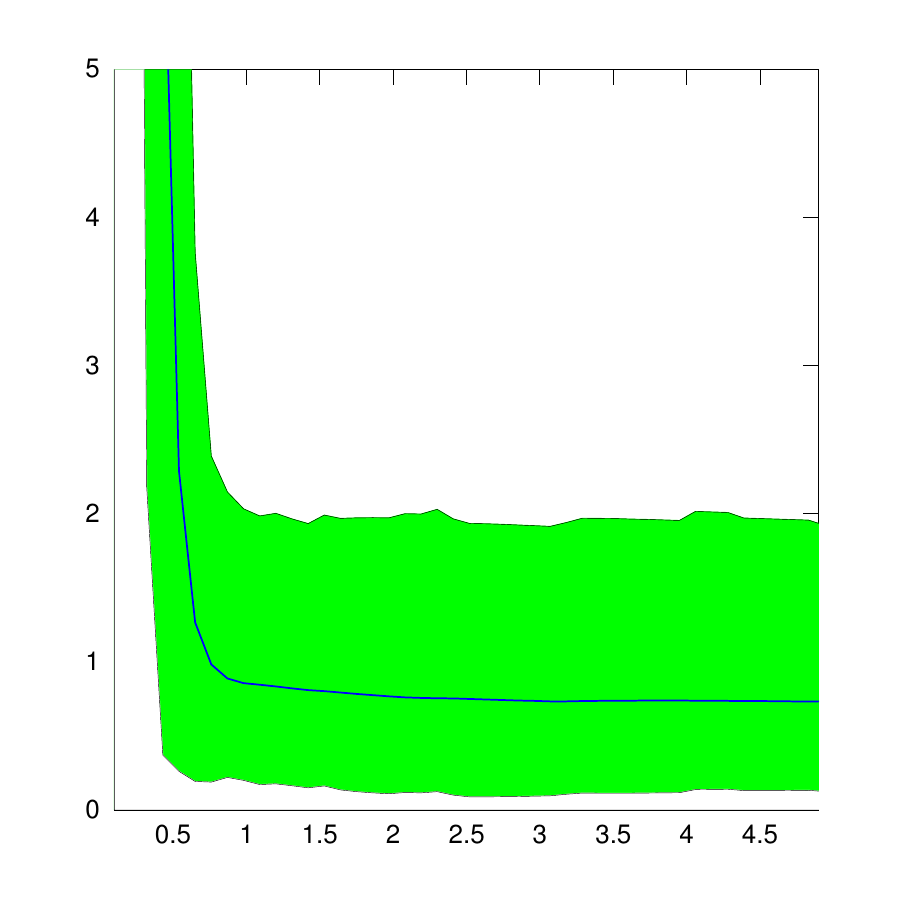}	 }
								\put(300,170){$\tilde{n}_s=0$}	\put(300,160){$\tilde{n}_b=1000$}
								\put(360,3){$\lambda_s$} \put(255,100){\rotatebox{90}{$Z_{\rm mom}$}}								
								
		\vspace{0.0cm}		
\end{picture}	
	\caption{ Expected significances $Z^{\rm std}$ (left) and $Z^{\rm mom}$ (right)  for the Rayleigh and exponential toy-models with background $\rho_b=2$, $\lambda_b=1$. The $Z^{\rm mom}$ includes the $8$ first moments (\ie~$P=8$).
The blue line is the expected value, the green areas correspond to one standard	deviation.
	 The $Z^{\rm mom}$ becomes unreliable when searching for a 'fat tail', for $\rho_s > 2.5 $, $\lambda_s < 0.8 $. 
		 	 \label{fig:signsZ}}
\end{figure}

It is instructive to first study the behaviour of our moment-based discovery test over the background-only pseudo-data. 
The expected $Z_{\rm std}$ and $Z_{\rm mom}$ with $P=8$  are shown in Fig.~\ref{fig:signsZ}, assuming a data sample of $1000$ events.
No  inconsistency  related to $Z_{\rm mom}$ appears when there is not fat tail. 
Note that we display the expected significances, so that the  fluctuations responsible of the look-elsewhere effect (LEE) \cite{LEE} do not appear.  The LEE can be  obtained in practice by evaluating the expected number of level-crossings. We check that the mean level-crossing number is roughly the same for $Z_{\rm std}$ and $Z_{\rm mom}$, on the interval where the signal is non-local, so that the LEE is expected to be approximatively the same for the two significances.
We observe that  $Z_{\rm mom}$ becomes not reliable when the hypothetical signal that one searches features a fat tail, \ie~ when $\rho_s$ ($\lambda_s$) is somewhat larger (smaller) than $\rho_b$ ($\lambda_b$). In that regime, $Z_{\rm mom}$ systematically grows large, detecting the existence of a signal while there is nothing to detect. The moment-based significance is thus totally wrong in that regime. This behaviour is common to the various toy-models with fat-tail signal we tested. It may be interesting to understand this behaviour in details, as this might open possibilities of corrections and thus extend the moment-based approach to the fat-tail case. We leave this exercise for a future work.
The moment-based approach is also expected to break down at small $\rho_s$ (large $\lambda_s$), where the signal becomes a localized bump. This is not obvious from Fig.~\ref{fig:signsZ}, but will appear in what follows. 


\begin{figure}
\begin{picture}(400,400)
\put(0,220){		\includegraphics[trim=0cm 0cm 0cm 0cm, clip=true,width=7.4cm]
		{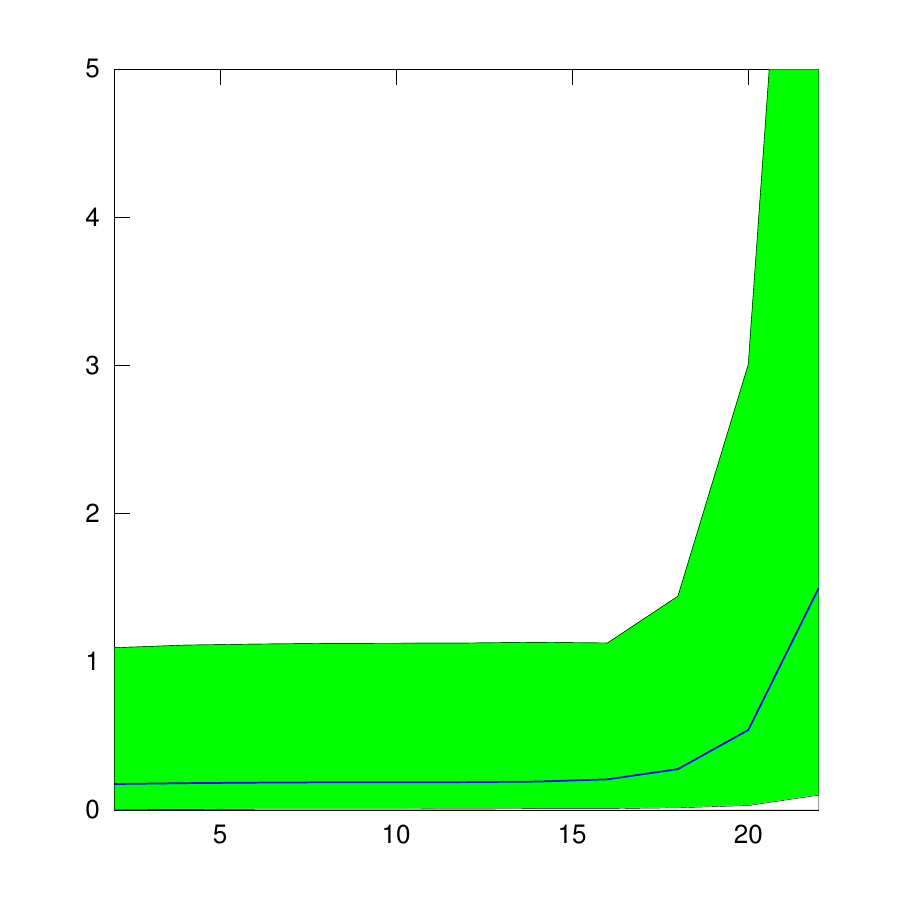}			}
				\put(50,390){$\tilde{n}_s=100$}	\put(50,380){$\tilde{n}_b=10000$}
				\put(110,225){$P$}
  						\put(253,300){\rotatebox{90}{$|Z_{\rm mom}-Z_{\rm std}|$}}
\put(250,220){					\includegraphics[trim=0cm 0cm 0cm 0cm, clip=true,width=7.4cm]
		{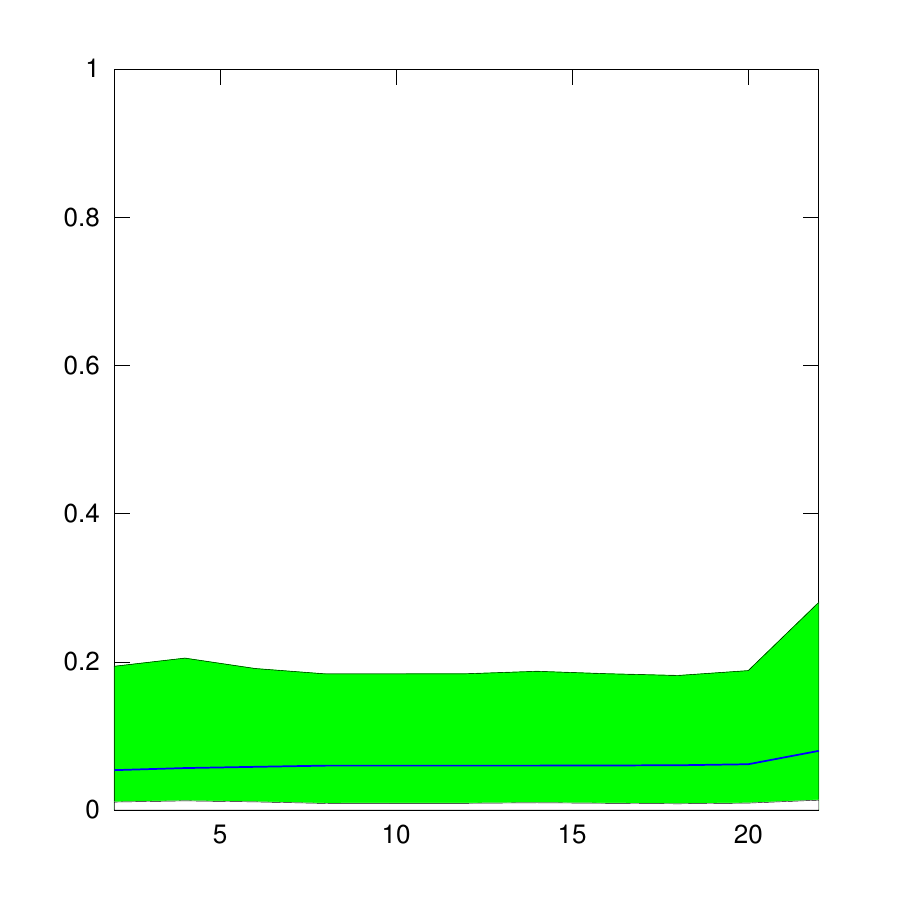}		}
						\put(300,390){$\tilde{n}_s=1000$}	\put(300,380){$\tilde{n}_b=100000$}
						\put(253,80){\rotatebox{90}{$|Z_{\rm mom}-Z_{\rm std}|$}}
										\put(360,225){$P$}
\put(0,0){							\includegraphics[trim=0cm 0cm 0cm 0cm, clip=true,width=7.4cm]
		{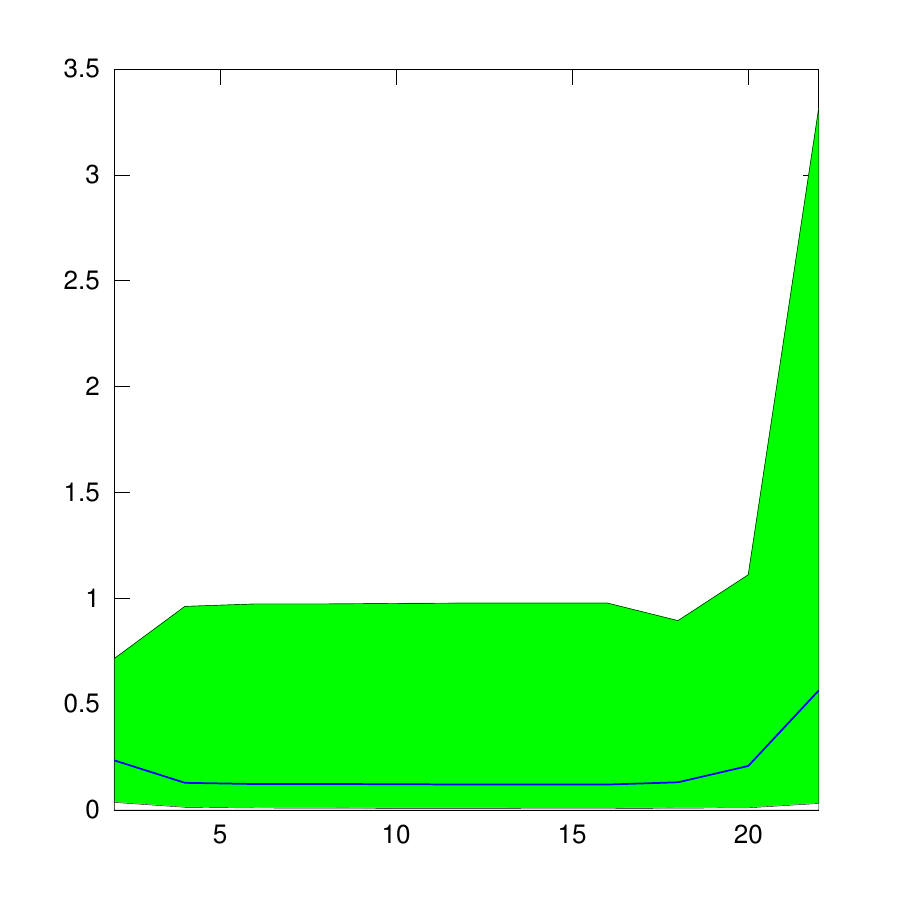}		}
						\put(50,170){$\tilde{n}_s=100$}	\put(50,160){$\tilde{n}_b=10000$}
						\put(3,300){\rotatebox{90}{$|Z_{\rm mom}-Z_{\rm std}|$}}
						\put(110,5){$P$}
\put(250,0){							\includegraphics[trim=0cm 0cm 0cm 0cm, clip=true,width=7.4cm]
		{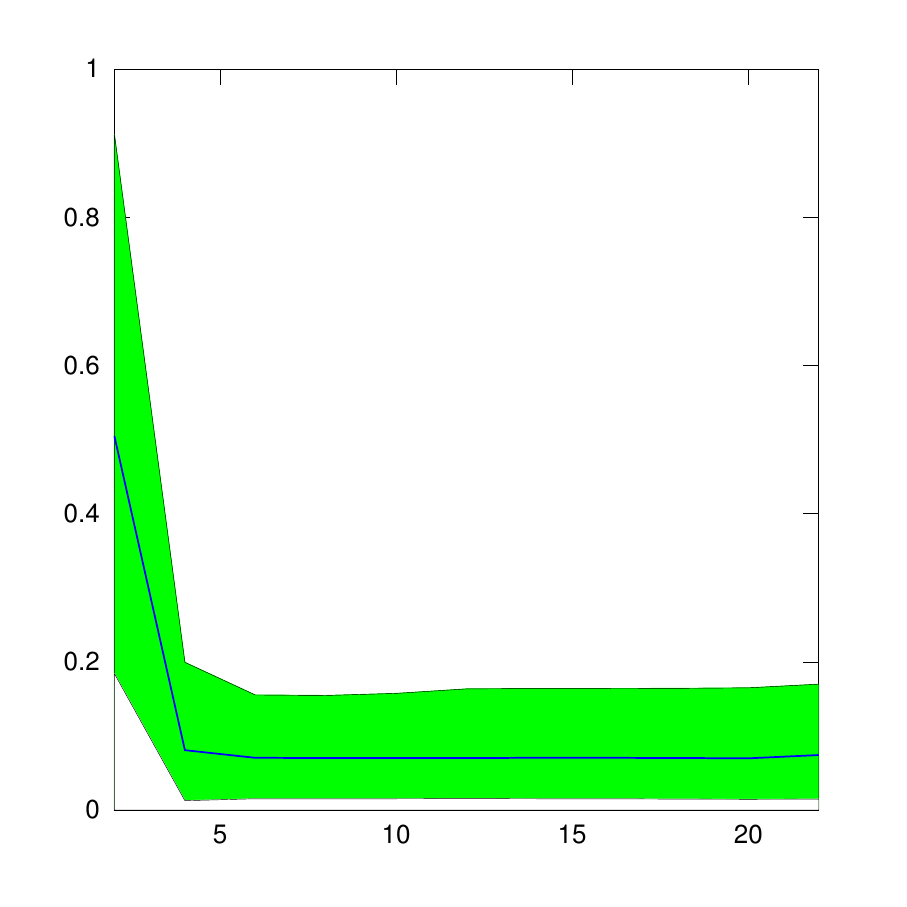}	}		
						\put(300,170){$\tilde{n}_s=1000$}	\put(300,160){$\tilde{n}_b=100000$}
						\put(3,80){\rotatebox{90}{$|Z_{\rm mom}-Z_{\rm std}|$}}
												\put(360,5){$P$}
		\end{picture}
		\vspace{0.0cm}
	\caption{ Examples of expected significance difference $|Z^{\rm mom}-Z^{\rm std}|$  for the Rayleigh toy-model with background $\rho_b=2$. Conventions are as in Fig. \ref{fig:signsZ}.
	 \textit{Top}: 	 
Signal with $\rho_s=1.5$ \textit{Bottom}: Signal with $\rho_s=2.5$ (fat tail).
		 \label{fig:signs}}
\end{figure}

\begin{figure}
\begin{picture}(400,200)
\put(0,0){			\includegraphics[trim=0cm 0cm 0cm 0cm, clip=true,width=7.4cm]
		{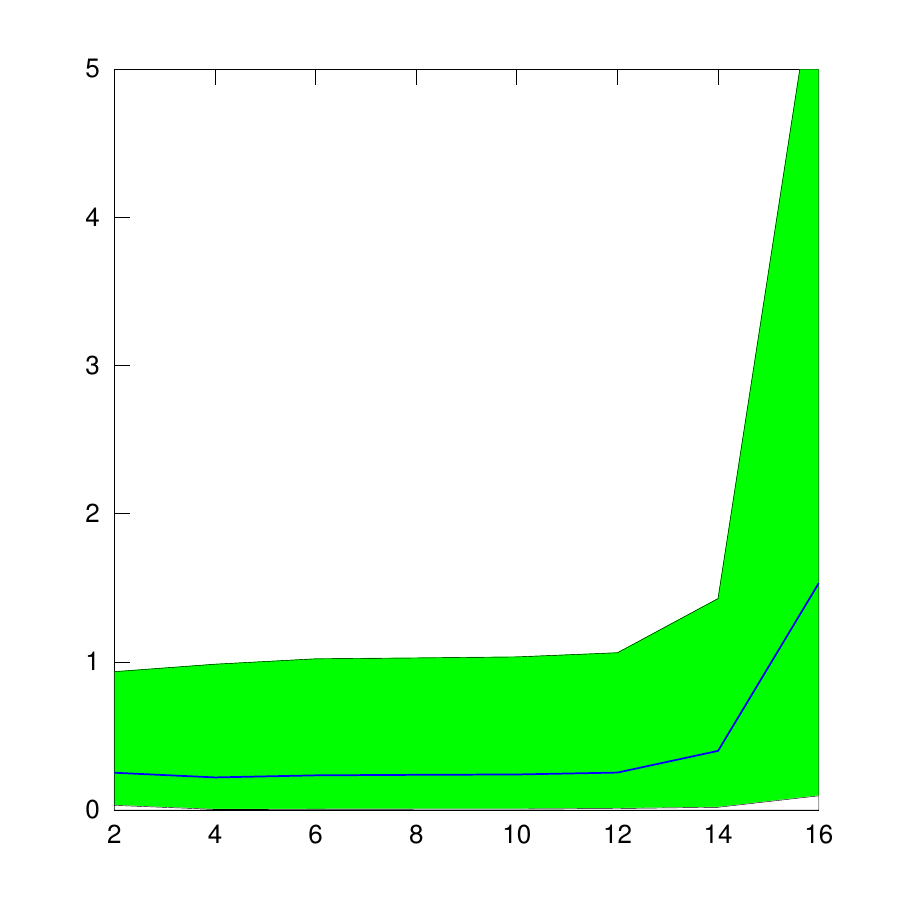}	}
							\put(50,170){$\tilde{n}_s=100$}	\put(50,160){$\tilde{n}_b=10000$}
							\put(110,5){$P$} 	\put(3,80){\rotatebox{90}{$|Z_{\rm mom}-Z_{\rm std}|$}}
	\put(250,0){							\includegraphics[trim=0cm 0cm 0cm 0cm, clip=true,width=7.4cm]
		{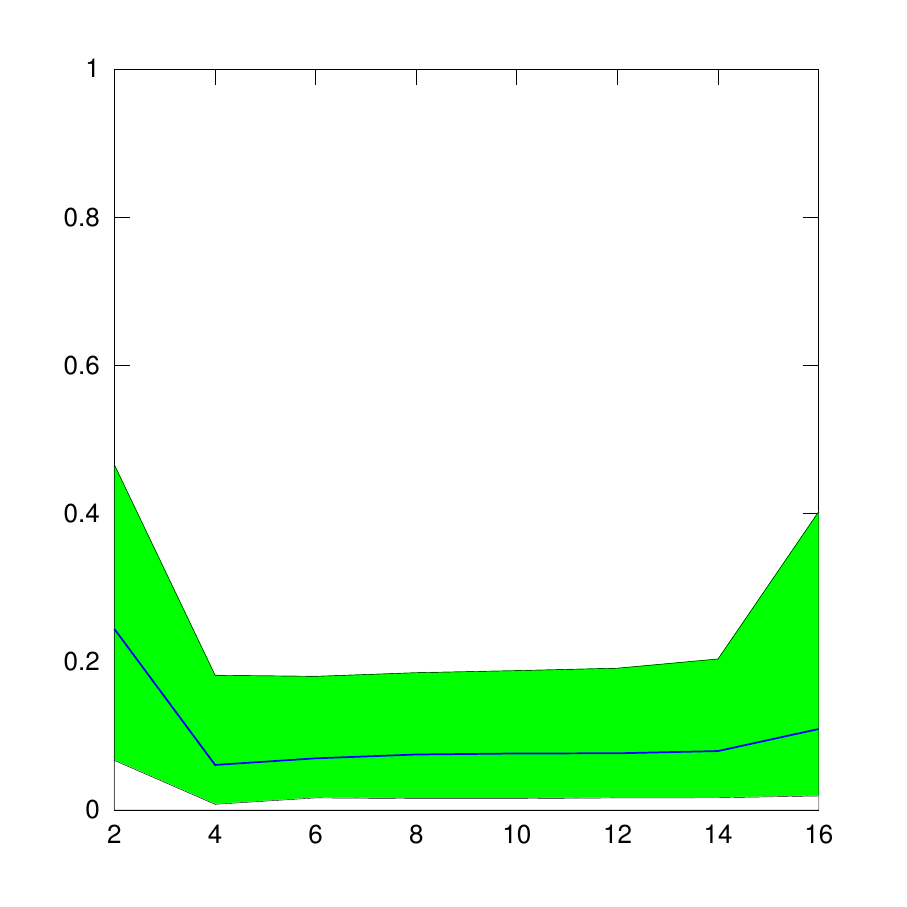}	 }
								\put(300,170){$\tilde{n}_s=1000$}	\put(300,160){$\tilde{n}_b=100000$}
								\put(360,5){$P$} 	\put(253,80){\rotatebox{90}{$|Z_{\rm mom}-Z_{\rm std}|$}}
		\vspace{0.0cm}		
\end{picture}	
	\caption{ Expected significance difference $|Z_{\rm mom}-Z_{\rm std}|$ (left) and moment-based significance $Z_{\rm mom}$ (right)  for the exponential toy-model with background $\lambda_b=1$, signal with $\lambda_s=2$. Conventions are as in Fig. \ref{fig:signsZ}.
	 	 \label{fig:signs2}}
\end{figure}

We can now focus on the domains of $\lambda$, $\rho$ where the  signal is non-local. This time we  introduce a signal into the pseudo-data,  such that $\tilde{n}_s/\tilde{n}_b=0.01$. Our aim is to check the validity of our (qualitative) claims about the equivalence between 
 $Z_{\rm std}$ and  $Z_{\rm mom}$. We compute the expected standard and moment-based significances from a large number of  pseudo-experiment and look at their difference, $|Z_{\rm std}-Z_{\rm mom}|$. As prescribed in Sec.~\ref{se:like_mom}, we plot the significance difference for the various values of $P$, until the moment covariance matrix becomes approximatively singular.

The Rayleigh and exponential toy-model, Figs. \ref{fig:signs},   \ref{fig:signs2} give similar conclusions. A \textit{plateau} appears over a large interval of $P$. In the examples considered, the significance difference is about $10\%$ of a standard deviation in average. The standard deviation on the difference does not go above $1\sigma$ for the case with few events, and is much smaller for the case with many events.
Beyond the examples displayed, one  observes that the mean significance difference and its  standard deviation   decrease with the sample size. 
Also, we observe that the significance remains stable with respect to the  total moment number $P$ over a sizeable range, enough to detect it without ambiguity. The interval of stability depends on the total event number. 
These observations confirm that  the moment-based likelihood matches rather well the standard likelihood when the condition of having a non-local  signal is fulfilled.  
We conclude that the standard likelihood can be safely replaced by the moment-based likelihood in that regime, with the technical benefits described in Sec.~\ref{se:limits}. 

\begin{figure}
\begin{picture}(400,250)
\put(0,0){				\includegraphics[trim=0cm 0cm 0cm 0cm, clip=true,width=7.4cm]
		{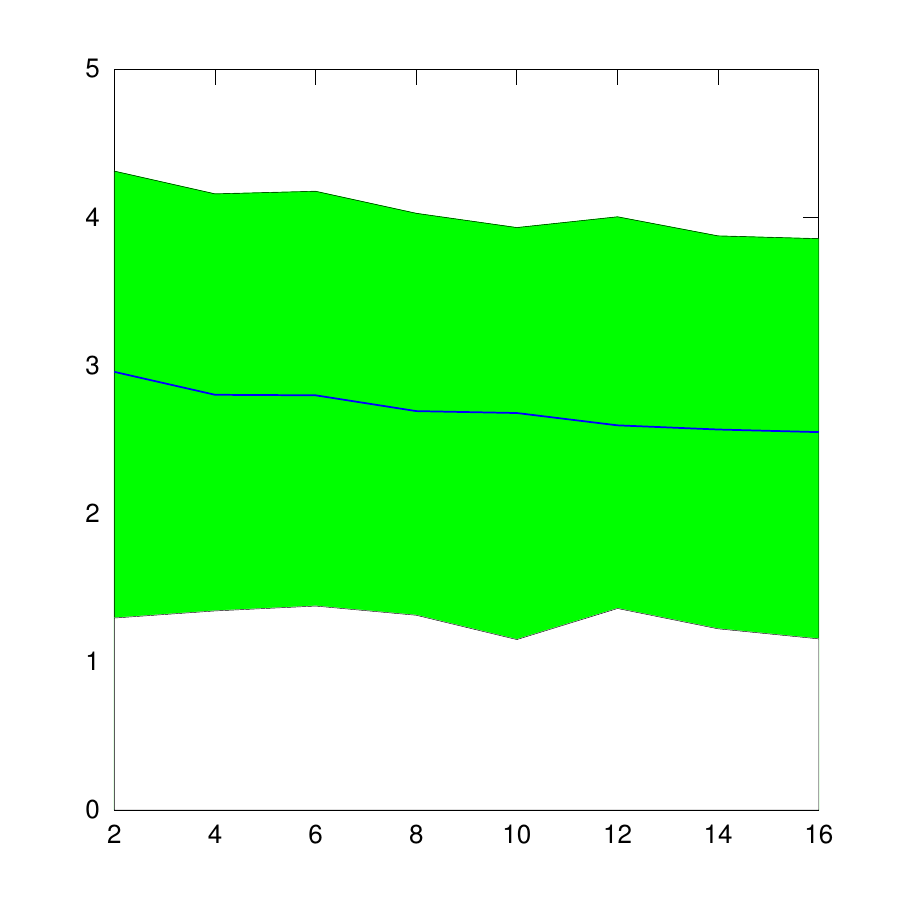}	}
							\put(50,170){$\tilde{n}_s=20$}	\put(50,160){$\tilde{n}_b=1000$} \put(110,5){$P$}
							\put(3,80){\rotatebox{90}{$|Z_{\rm mom}-Z_{\rm std}|$}}
\put(250,0){				\includegraphics[trim=0cm 0cm 0cm 0cm, clip=true,width=7.4cm]
		{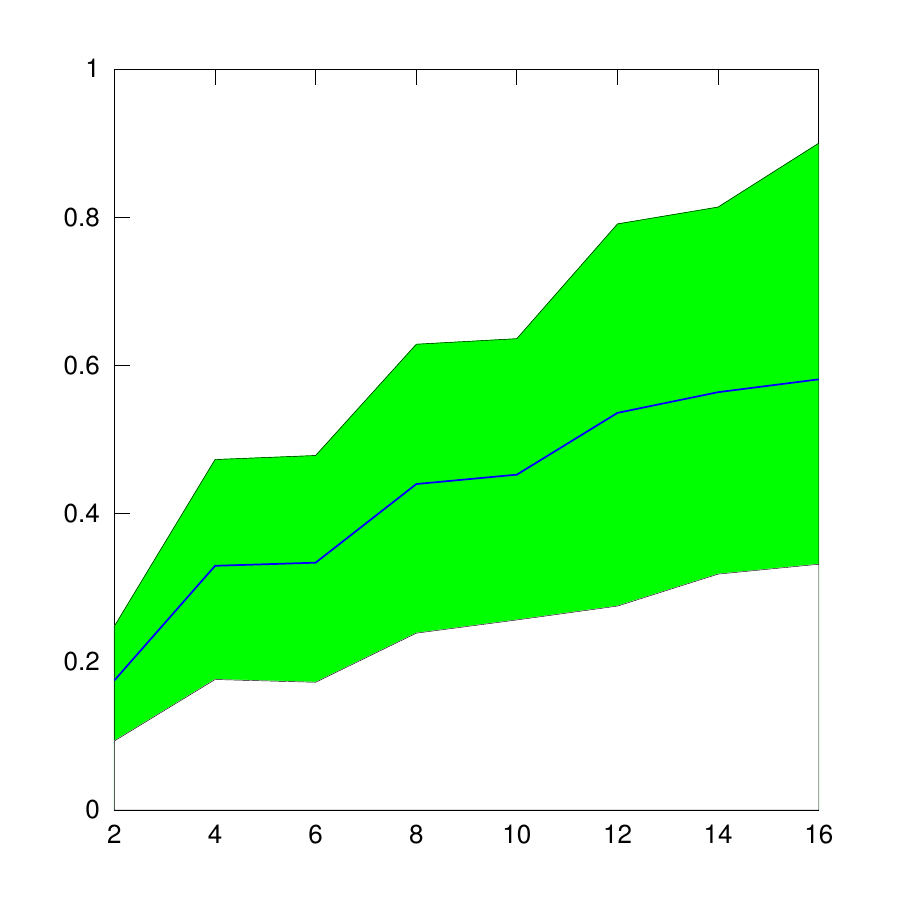}	}
										\put(300,170){$\tilde{n}_s=20$}	\put(300,160){$\tilde{n}_b=1000$}	\put(360,5){$P$}													\put(253,80){\rotatebox{90}{$Z_{\rm mom}$}}
		\end{picture}
	\caption{ Examples of expected significance difference $|Z_{\rm mom}-Z_{\rm std}|$  for the Rayleigh toy-model with background $\rho_b=2$ and a Gaussian bump at $x=3$ with width $\sigma=0.03$.  Conventions are as in Fig. \ref{fig:signsZ}.	
		 \label{fig:signs3}}
\end{figure}
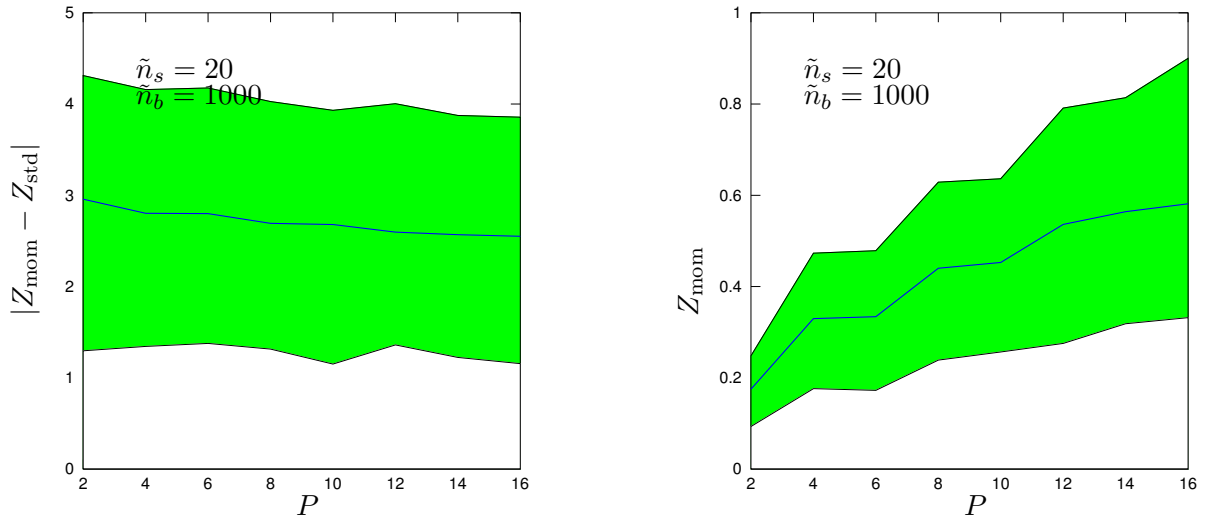

In contrast, in presence of a thin bump, Fig. \ref{fig:signs3},  we see that no \textit{plateau} appears. Rather, $Z_{\rm mom}$ grows slowly with $P$, until the covariance becomes singular. This corresponds to the case described in Sec.~\ref{se:like_mom} where the information is not mostly contained in the first moments. The behaviour exemplified in  Fig. \ref{fig:signs3} is general to any peaked signal in our toy-models, including the $\rho_s\ll \rho_b$ and  $\lambda_s\gg \lambda_b$ cases. Again, it might be possible to characterize more precisely this behaviour. We leave this for further study. In addition of being unstable, $Z_{\rm mom}$ is much smaller than $Z_{\rm std}$. The moment-based likelihood seems therefore inappropriate for peaked signals, as expected from general arguments.

%
%
%
%
%
%
%
%
%
%

 \newpage

\section{Conclusions and outlook}  \label{se:summary}

The analysis of the shape of a data sample is an  exercise frequently encountered in experimental physics. 
Among many topics, it plays an important role in the searches for a new physics signal in high-energy data, like the ones collected at the LHC. 

In this work, we introduce a new kind of likelihood based on the moments of data distributions. Both binned and unbinned cases are treated. The multivariate case is also derived,
 and leads to a tensor-variate normal likelihood. 
A review of the standard likelihoods is included, and simplified statistical tests are also provided,  such that the paper is self-contained from the perspective of shape analysis. A particular focus is put on the shape analysis for  signal searches.

It appears that the moment-based likelihoods, whenever they can compete with the standard ones, can  simplify the tasks related to signal searches commonly encountered in high-energy physics.  
 The key point is that the hypothetical distribution of the background often needs to be estimated from a fit of Monte-Carlo simulations. This step of fitting is rather tricky as it can 
easily introduce small deviations from the true hypothetical background, that can be misinterpreted with the presence of a signal. This fit problem increases drastically for multivariate shape analysis, \ie~when several observables are treated at the same time. 
The moment-based likelihoods totally bypass the  step of fitting, as the moments are trivially deduced from the MC simulations, and the MC error stays well under control. 


Our moment-based approach is promising for the searches for  non-local signals, where most of the information is contained in the first moments. Note the effective operators that enclose the low energy effects of new physics typically produce such non-local signals. Instead, when the signal is localized over the background (like a ``bump''), the moment-based likelihood cannot be as efficient as the standard likelihood.  This case of a local signal is familiar and carefully treated in high-energy physics, such that the standard and moment-based approaches are complementary.


We exemplify and check the  moment-based approach by computing discovery tests within  toy-models representative of new physics searches at the LHC. It appears that the standard and moment-based significances are in good agreement when the signal is non-local. 
We also observe that the behaviour of the moment-based significance in presence of a fat-tailed signal constantly fails in a similar way. This pattern is rather striking, and would deserve more investigation, that we leave as an interesting open issue.

Apart from the ``fat tail'' issue, further formal developments would be certainly useful to define more precisely the conditions of equivalence between moment-based and standard likelihoods. 
We hope that this work  opens a useful set of possibilities for further development and improvement of shape analysis and  signal searches techniques.

\section*{Acknowledgements} 

The author would like to thank Veronica Sanz and Glen Cowan for fruitful discussions, and Thomas Kloss and Raissa Estrela
 for reading the manuscript.  The author acknowledges the Brazilian Ministry of Science, Technology and Innovation for financial support, and the  Les Houches 2013 PhysTeV workshop  where a part of this work was initiated.
 
\vspace*{2cm}

\appendix

\noindent{\Large\bf Appendix}

\section{Likelihoods for orthonormal decompositions \label{app:ortho}}

It is worth mentioning the alternative possibility of a decomposition over an orthonormal basis.
We will not follow this route because the success of the approach might be  more problem-dependent, while the moment decomposition is fairly  universal. 

Starting  from the decomposition Eq. \ref{eq:decomp}, we can use the orthonomality relation. 
 In general $\left<g_p,g_q\right>=\int_\mathcal{D}dx\, w(x) g_p(x) g_q(x)=\delta_{pq}$, where $w(x)$ is a specific weight function. 
 The coefficients are then determined as 
\be
a_p=\int_\mathcal{D}dx\, w(x) g_p(x) f_X(x)\,.
\ee
Given $n$ events, an estimator of $a_p$ is given by
\be
\hat{a}_p=\frac{1}{n}\sum_i^n w(X_i)g_p(X_i)\,,
\ee
such that $E[\hat{a}_p]= a_p$. 
By the Central Limit Theorem (CLT), for large number of events $n$, each of the $a_p$ coefficients follow a normally distributed law with mean $a_p$. 
 The $a_p$ being estimated from the same data, they are correlated and described by a multivariate normal. Their covariance matrix $\Sigma$ is estimated by 
\be
\hat{\Sigma}= \frac{1}{n}\left(\sum_i^n w^2(X_i) g_p(X_i) g_q(Xi) - \hat{a}_p\hat{a}_q\right)\,.
\ee
Note a simplification occurs in case of a Fourier series, as $w=1$, $g_pg_q=g_{p+q}$, such that 
\be
\hat{\Sigma}_{pq}= \frac{1}{n}\bigg(\hat{a}_{p+q} - \hat{a}_p\,\hat{a}_q\bigg)\,.
\ee
The choice of an appropriate basis would depend to some extent on the shape of $f_X$. 
For example, the Fourier expansion would certainly be appropriate for angular variables distributions.  
We will however not follow these possibilities as they might be problem-dependent.

\section{ Densities and moments for the pseudo-data  \label{app:distributions}}

The Rayleigh and exponential PDFs are respectively given as \be
f_X(x)=\frac{x}{\rho^2}e^{-x^2/2\rho^2}\,,\quad f_X(x)=\lambda e^{-\lambda x}\,.
\ee
Their respective raw moments are 
\be
m_p=(\sqrt{2}\rho)^p \Gamma(1+\frac{p}{2})\,,\quad m_p\frac{p!}{\lambda^p}
\ee
The raw moments of the normal PDF $\exp(-\frac{(x-\mu)^2}{2\sigma^2} )$ are given by 
\be
\sigma^p  (-i\sqrt 2)^p  U\left( -\frac{p}{2},\frac{1}{2},-\frac{\mu^2}{2\sigma^2} \right)\,,
\ee
where $U$ is the confluent hypergeometric of the second kind
\be
U(a,b,z)=\frac{\pi}{\sin(b\pi)} \left(\frac{_1F_1(a;b;z)}{\Gamma(1+a-b) \Gamma(b)}  -z^{1-b} \frac{_1F_1(1+a-b;2-b;z)}{\Gamma(a)\Gamma(2-b) } \right).
\ee


\begin{thebibliography}{99}

\bibitem{Dumont:2013wma} 
  B.~Dumont, S.~Fichet and G.~von Gersdorff,
  ``A Bayesian view of the Higgs sector with higher dimensional operators,''
  JHEP {\bf 1307}, 065 (2013)
  [arXiv:1304.3369 [hep-ph]].
  
  
  
\bibitem{Ellis:2014dva} 
  J.~Ellis, V.~Sanz and T.~You,
  ``Complete Higgs Sector Constraints on Dimension-6 Operators,''
  arXiv:1404.3667 [hep-ph].


\bibitem{GMM} 
  A.R.~Hall,
  ``Generalized Method of Moments,''
  Advanced Texts in Econometrics,
   Oxford University Press.

\bibitem{lehmann}
E.L.~Lehmann, G.~Casella,
``Theory of point estimation'',
Springer texts in statistics, 1998, Springer.




\bibitem{Muirhead}
R.J.~Muirhead,
`` Aspects of multivariate statistical theory,''
 Wiley series in probability and mathematical statistics. John Wiley, 1982. 

%
%


\bibitem{Grzadkowski:2010es} 
  B.~Grzadkowski, M.~Iskrzynski, M.~Misiak and J.~Rosiek,
  ``Dimension-Six Terms in the Standard Model Lagrangian,''
  JHEP {\bf 1010}, 085 (2010)
  [arXiv:1008.4884 [hep-ph]].

\bibitem{Buchmuller:1985jz} 
  W.~Buchmuller and D.~Wyler,
  ``Effective Lagrangian Analysis of New Interactions and Flavor Conservation,''
  Nucl.\ Phys.\ B {\bf 268}, 621 (1986).

\bibitem{Masso:2014xra} 
  E.~Masso,
  ``An Effective Guide to Beyond the Standard Model Physics,''
  arXiv:1406.6376 [hep-ph].


\bibitem{NP_scale} 
  S.~Fichet,
  ``Probing the scale of New Physics at the LHC: The example of Higgs data,''
  Nucl.\ Phys.\ B {\bf 884}, 379 (2014)
  [arXiv:1307.0544 [hep-ph]].

\bibitem{Chen:2014xra} 
  C.~-R.~Chen and I.~Low,
  ``A Double Take on New Physics in Double Higgs Production,''
  arXiv:1405.7040 [hep-ph].

\bibitem{Wilk} 
 S.~ Wilks,
 `` The Large-Sample Distribution of the Likelihood Ratio for Testing Composite Hypotheses,''
  The Annals of Mathematical Statistics 9 (1938), no. 1, 60--62
  
  \bibitem{Wald} 
A.~ Wald,`` Tests of Statistical Hypotheses Concerning Several Parameters When the
Number of Observations is Large, ''
Transactions of the American Mathematical Society,
Vol. 54, No. 3 (Nov., 1943), pp. 426-482.
  
  
   
\bibitem{LEE} 
E.~Gross and O.~Vitells, 
`` Trial factors for the look elsewhere effect in high energy physics,''
  	Eur.Phys.J.C70:525-530, 2010

\bibitem{asimov}

G.~ Cowan, K.~ Cranmer, E.~ Gross, O.~ Vitells,
``Asymptotic formulae for likelihood-based tests of new physics,''
Eur.Phys.J.C71:1554,2011




\end{thebibliography}
\end{document}